\documentclass[12pt, oneside, letterpaper]{revtex4-2}   
 \usepackage[utf8]{inputenc}
\usepackage{mwe,float}
\usepackage{caption,subcaption, url}
\usepackage{natbib}
\usepackage[nointegrals]{wasysym}
\usepackage{rotating}
\usepackage[pdftex]{color}
\usepackage{array,tabu,multirow}
\usepackage{graphicx,bm}
 \usepackage{floatpag} 
 \rotfloatpagestyle{empty}
\usepackage{amsmath}
\usepackage{amsthm}
\usepackage{amssymb} 
\usepackage{latexsym,upgreek}
\usepackage{hyperref}
\usepackage{rotating}
\usepackage{xpatch,bigints}
\usepackage{booktabs,subcaption}
\makeatletter
\AtBeginDocument{\xpatchcmd{\@thm}{\thm@headpunct{.}}{\thm@headpunct{\enskip}}{}{}}
\makeatother 
\usepackage{gensymb}
\usepackage{mathdots}

\newcommand{\qand}{\quad \mbox{and} \quad}

\newcommand{\tbs}[1]{\textbf{\textsf{#1}}}
\usepackage[version=4]{mhchem}

\usepackage{graphicx,kecpm,slashed}
\usepackage{makeidx}
\makeindex

\DeclareMathOperator{\arcsinh}{arcsinh}

\newcommand{\bex}{\begin{example}}
\newcommand{\eex}{\end{example}}
\newcommand{\besp}{\begin{split}}
\newcommand{\ensp}{\end{split}}

\renewcommand{\thefootnote}{\fnsymbol{footnote}}

\newcommand{\by}{\times}

\newcommand{\ovl}{\overline}

\newcommand{\bos}{\boldsymbol}
\newcommand{\tbf}{\textbf}

\newcommand{\btab}{\begin{tabular}}
\newcommand{\etab}{\end{tabular}}
\newcommand{\barr}{\begin{array}}
\newcommand{\earr}{\end{array}}
\newcommand{\bpm}{\begin{pmatrix}}
\newcommand{\epm}{\end{pmatrix}}
\newcommand{\bit}{\begin{itemize}}
\newcommand{\eit}{\end{itemize}}
\newcommand{\ben}{\begin{enumerate}}
\newcommand{\een}{\end{enumerate}}
\newcommand{\bct}{\begin{center}}
\newcommand{\ect}{\end{center}}

\newcommand{\bes}{\begin{split}}
\newcommand{\ens}{\end{split}}

\newcommand{\lt}{\left}
\newcommand{\rt}{\right}

\begin{document}
\title{Flat Space, Dark Energy, and the 
Cosmic Microwave Background}

\author{Kevin Cahill}
\affiliation{Department of Physics and Astronomy\\
University of New Mexico\\
Albuquerque, New Mexico 87106}
\date{\today}

\begin{abstract}
This paper reviews some of the results of the Planck collaboration and shows how to compute the distance from the surface of last scattering, the distance from the farthest object that will ever be observed, and the maximum radius of a density fluctuation in the plasma of the CMB.  It then explains how these distances together with well-known astronomical facts imply that space is flat or nearly flat and that dark energy is 69\% of the energy of the universe.
\end{abstract}

\maketitle

\section{Cosmic Microwave Background Radiation
\label{Cosmic Microwave Background Radiation sec}}

The cosmic microwave background (CMB)
was predicted by Gamow
in 1948~\citep{Gamow:1948oesg}, 
estimated to be at a temperature of 5\,K by
Alpher and Herman in 1950~\citep{Alpher:1950zz},
and discovered by Penzias and Wilson
in 1965~\citep{Penzias:1965wn}\@.
It has been observed in increasing detail 
by Roll and Wilkinson 
in 1966~\citep{Roll:1966zz}, 
by the Cosmic Background Explorer (COBE) 
collaboration in 
1989--1993~\citep{Mather:1991pc, Smoot:1992td},
by the 
Wilkinson Microwave Anisotropy Probe 
(WMAP) collaboration in 
2001--2013~\citep{Peiris:2003ff, Bennett:2012zja},
and by the Planck collaboration in
2009--2019~\citep{Ade:2013zuv, Ade:2015xua, Akrami:2018vks, Aghanim:2018eyx}\@.
\par
\begin{figure}
\centering
\tbs{Temperature fluctuations of the cosmic microwave background radiation}
\par\medskip
\includegraphics[trim={1.7in 0.2in 1.7in 0.15in, clip}, 
width=0.65\textwidth]{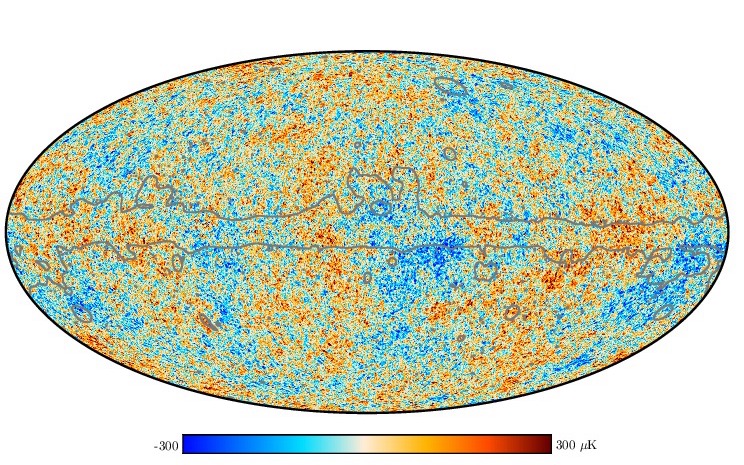}
\caption{CMB temperature fluctuations 
over the celestial sphere
as measured by the Planck satellite.  
The average temperature is 2.7255 K\@. 
The gray line outlines our galaxy.
(arXiv:1807.06205 [astro-ph.CO], A\&A
doi.org/10.1051/0004-6361/201833880)}
\label {CMB face of god}
\end{figure}

 
The Planck collaboration 
measured CMB radiation 
at nine frequencies from 30 to 857 GHz
by using a satellite
at the Lagrange point L$_2$
in the Earth's shadow some
1.5$\by 10^6$ km farther
from the Sun~\citep{Akrami:2018vks, Aghanim:2018eyx}\@.
Their plot of the temperature $T(\theta,\phi)$ 
of the CMB radiation 
as a function of the angles
$\theta$ and $\phi$ in the sky 
is shown in Fig.~\ref{CMB face of god}
with our galaxy outlined in gray.
The CMB radiation is that of
a 3000 K blackbody redshifted to 
$T_0 = 2.7255 \pm 0.0006$K~\citep{Fixsen2009}\@.
After correction for the motion of the Earth,
the temperature of the CMB 
is the same
in all directions apart from
anisotropies of $ 300 \, \upmu$K
shown in red and blue.
The CMB photons 
have streamed freely 
since the baryon-electron plasma
cooled to 3000 K making
hydrogen atoms stable
and the plasma transparent.
This time of initial transparency,
some 380,000 years
after the big bang,
is called decoupling or
recombination.
\par
\begin{figure}
\centering
\tbs{Polarization of the cosmic microwave background radiation}
\par\bigskip
\includegraphics[angle=90, 
width=0.90\textwidth]{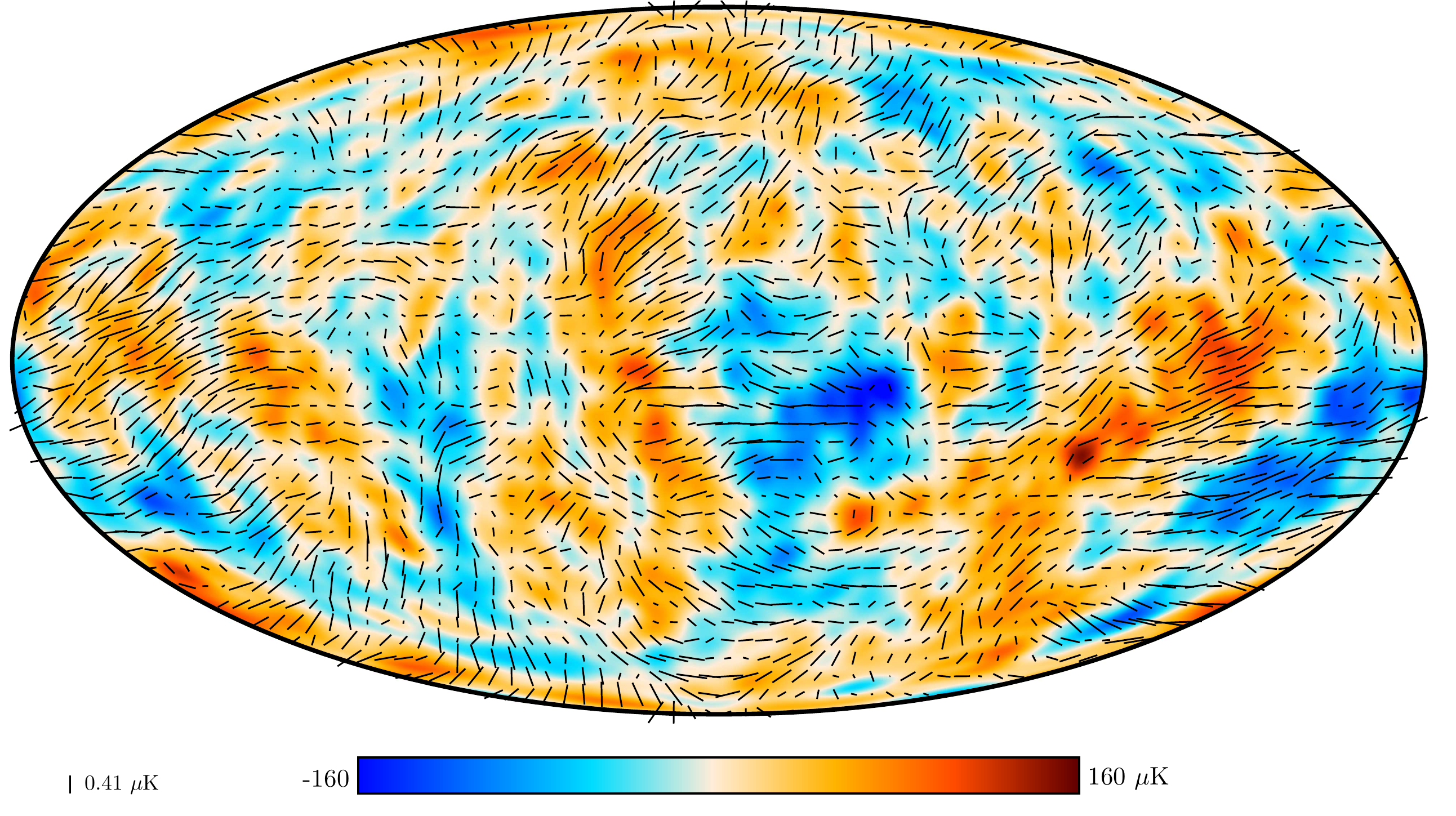}
\caption{The polarization field
superimposed upon the temperature map.
(arXiv:1807.06205 [astro-ph.CO], A\&A
doi.org/10.1051/0004-6361/201833880)}
\label {polarization fig}
\end{figure}
\begin{figure}
\centering
\tbs{The lensing map}
\par\bigskip
\includegraphics[
width=0.90\textwidth]{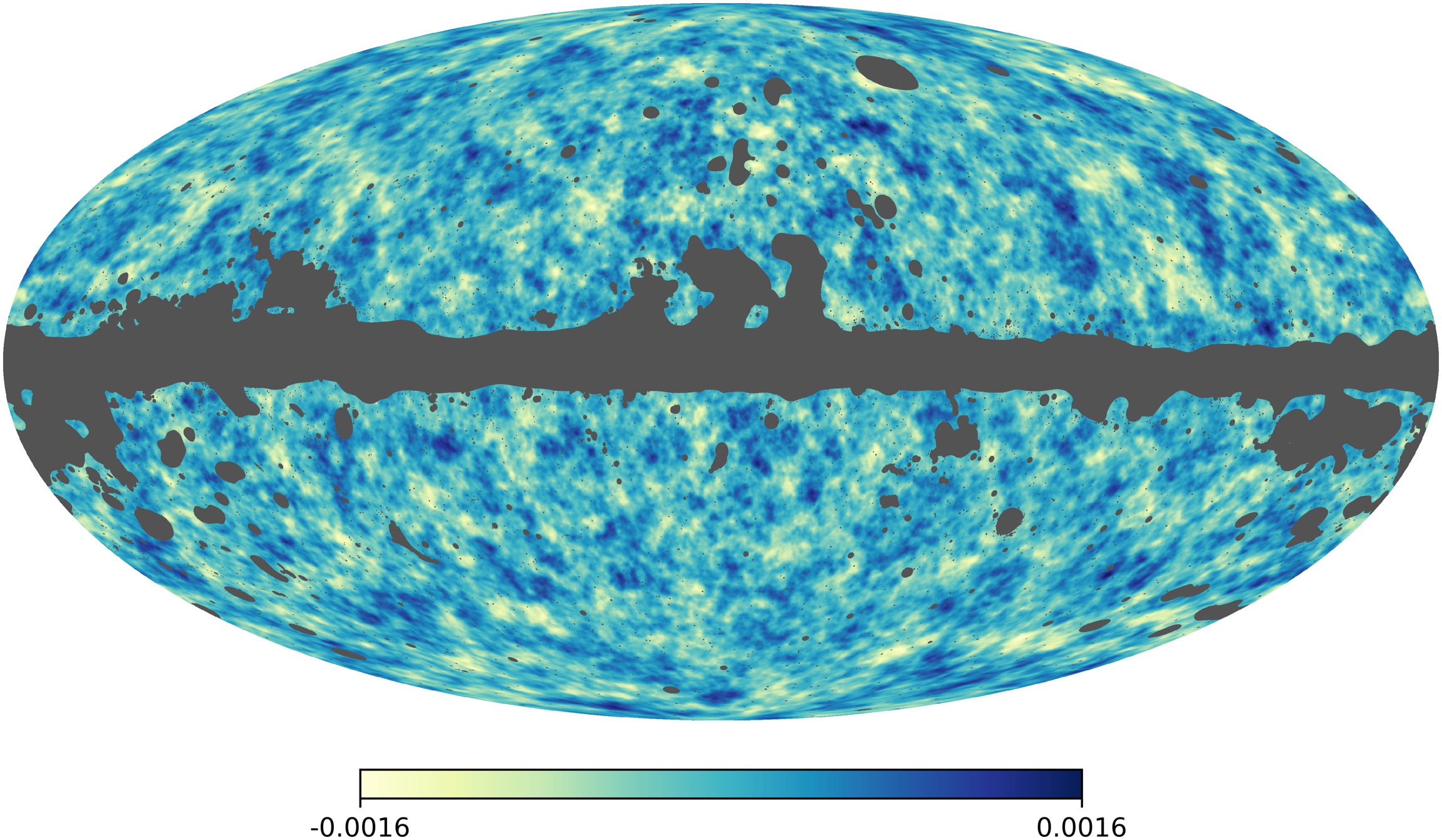}
\caption{The gravitational potential between Earth and the surface of last scattering.
(arXiv:1807.06205 [astro-ph.CO], A\&A 
doi.org/10.1051/0004-6361/201833880)}
\label {lensing fig}
\end{figure}

The CMB photons are polarized
because they have scattered
off electrons in the 
baryon-electron plasma
before recombination
and off electrons
from interstellar hydrogen
ionized by radiation from stars.
The Planck collaboration 
measured the
polarization of the CMB photons
and displayed it
in a graph reproduced
in Fig.~\ref{polarization fig}\@.
They also used
gravitational lensing
to estimate the 
gravitational potential
between Earth and
the surface of last scattering.
Their lensing map is shown
in Fig.~\ref{lensing fig}
with our galaxy in black.

\begin{figure}[htbp]
\centering
\tbs{Theoretical fit to the temperature anisotropies}
\par
\includegraphics[width=\textwidth]{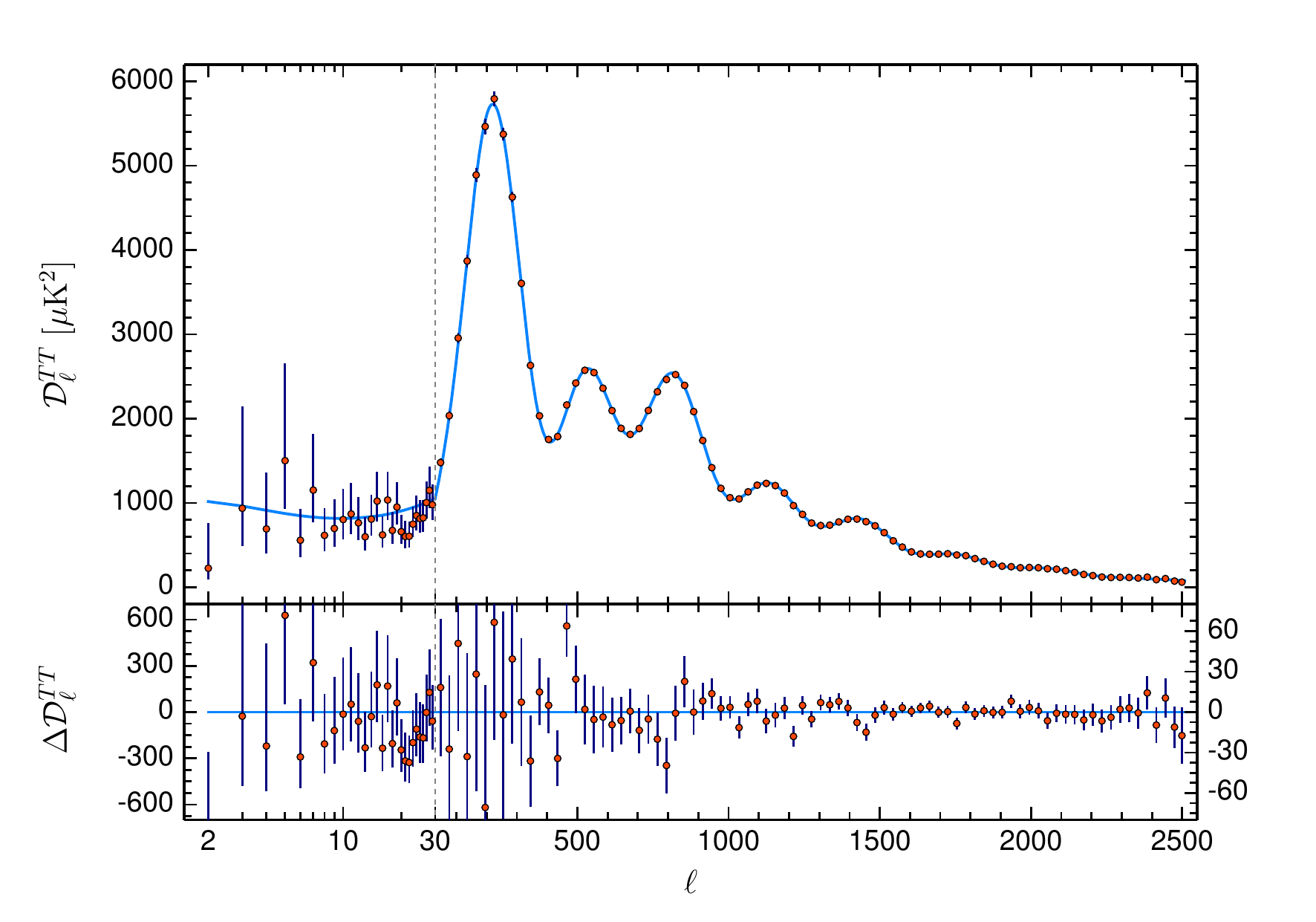}
\caption{The temperature-temperature 
(TT) power spectrum 
$\mathcal{D}^{TT}_\ell = \ell(\ell+1) C_\ell/2\pi$ 
with its residuals in the lower panel
is plotted against the multipole moment
$\ell$ on a scale that goes 
from logarithmic to linear at $\ell=30$\@.
The blue curve is the 
six-parameter $\Lambda$CDM fit.
(Planck collaboration, arXiv:1807.06209)}
\label {CMB fit}
\end{figure} 

\par
The Planck collaboration expanded
the temperature $T(\theta,\phi)$
they measured
in spherical harmonics
\begin{equation}
T(\theta,\phi) = \sum_{\ell = 0}^\infty
\sum_{m=-\ell}^\ell a_{\ell,m} 
\, Y_{\ell, m}(\theta,\phi).
\label {T expansion}
\end{equation}
They used the coefficients 
\begin{equation}
a_{\ell,m} = \int_0^{2\pi} \!\!d\phi \int_0^\pi \!\! 
\sin\theta \, d\theta \,\,\,
Y^*_{\ell, m}(\theta,\phi) \, T(\theta,\phi)
\label {alm =}
\end{equation}
to define a
temperature-temperature (TT)  
power spectrum  
\begin{equation}
\mathcal{D}^{TT}_\ell = {}
\frac{\ell(\ell + 1)}{2\pi(2\ell + 1)} 
\sum_{m=-\ell}^\ell |a_{\ell,m}|^2 .
\label {T power spectrum}
\end{equation}
They similarly
represented their
measurements of CMB polarization 
and gravitational lensing
as a temperature-polarization
(TE) power spectrum 
$\mathcal{D}^{TE}_\ell$,
a polarization-polarization 
(EE) power spectrum 
$\mathcal{D}^{EE}_\ell$,
and a lensing spectrum $C^{\phi\phi}_\ell$\@.
They were able to fit a simple, flat-space
model of a universe with cold dark matter
and a cosmological constant  
to their TT, TE, EE, and 
lensing data
by using only six parameters.
Their amazing 
fit to the TT spectrum
is the blue curve
plotted in Fig.~\ref{CMB fit}\@.
\par
The temperature-temperature power spectrum
$\mathcal{D}^{TT}_\ell$
plotted in Fig.~\ref{CMB fit}
is a snapshot
at the moment of initial transparency
of the temperature distribution
of the rapidly expanding plasma of 
dark matter,
baryons, electrons, neutrinos, and photons
undergoing tiny ($ 2\by 10^{-4} $) 
acoustic oscillations.
In these oscillations,
gravity opposes radiation pressure, and  
$ |\Delta T(\theta,\phi) | $ 
is maximal
both when the oscillations are most compressed and 
when they are most rarefied.
Regions that gravity has squeezed to
maximum compression at transparency
form the first and highest peak.
Regions that have bounced off their first 
maximal compression and that 
have expanded under
radiation pressure 
to minimum density at transparency
form the second peak.
Those at their second maximum compression at transparency
form the third peak, and so forth.
Decoupling was not instantaneous:
the fractional ionization of hydrogen
dropped from 0.236 at 334,600
years after the big bang
to 0.0270 at 126,000 years 
later~\citep[p. 124]{Weinberg2010}\@. 
The rapid high $\ell$ oscillations
are out of phase with each other
and are diminished.
\par
The Planck collaboration 
found their data to be consistent 
with a universe in which space is
flat but expanding
due to the energy density of 
empty space (dark energy
represented by a 
cosmological constant $\Lambda$)\@.
In their model, 84\% of the matter 
consists of invisible particles 
(dark matter) that 
were cold enough to be nonrelativistic 
when the universe had 
cooled to $T \sim 10^6\,$K 
or $kT \sim 100$ eV\@.
The Planck collaboration were able
to fit their $\Lambda$-cold-dark-matter 
($\Lambda$CDM) model 
as shown in Fig.~\ref{CMB fit}
to their huge sets of data 
illustrated by 
Figs.~\ref{CMB face of god}--\ref{lensing fig}
by adjusting only
six cosmological parameters.
In so  doing, they determined 
the values of these six quantities:
the present baryon density $\rho_{b 0}$
(including nuclei and electrons), 
the present cold-dark-matter 
density $\rho_{d 0}$,
the angle $\theta_s$ subtended in the sky
by disks whose radius is the sound horizon
at recombination,
the optical depth $\tau$ between Earth and the
surface of last scattering,
the amplitude $A_s$ of the fluctuations seen in
Figs.~\ref{CMB face of god}--\ref{CMB fit}, and 
the way $n_s$ in which fluctuations
vary with their wavelength.
Their estimates are~\citep[col. 7, p. 15]{Aghanim:2018eyx} 
\begin{equation}
\begin{array}{ll}
\rho_{b 0} = (4.211 \pm 0.026)\by 10^{-28} \, \mbox{kg/m}^3
\quad 
& \rho_{d 0} = (2.241 \pm 0.017) \by 10^{-27} \, \mbox{kg/m}^3 \quad
\\
\theta_s = 0.0104101 \pm 0.0000029  
&
\tau = 0.0561 \pm 0.0071
\\
\ln\lt(10^{10}A_s\rt) = 3.047 \pm 0.014
&  n_s = 0.9665 \pm 0.0038 .
\end{array}
\label {the six parameters}
\end{equation}
They also estimated 
some 20 other 
cosmological parameters~\cite[col. 7, p. 15]{Aghanim:2018eyx}  
and established 
flat $\Lambda$CDM
as the standard model of cosmology.
\par
Section~\ref{The Standard Model of Cosmology sec}
explains comoving coordinates,
Friedmann's equation,
the critical density, and some
basic cosmology.
Section~\ref{How the Scale Factor Evolves sec}
explains how the scale factor $a(t)$,
the redshift $z$,
and the densities
of matter, radiation, and empty space 
evolve with time 
and computes
the comoving distance $r_d$
from the surface of decoupling
or last scattering
and the comoving distance
from the most distant object
that will ever be observed.
Section~\ref{The Sound Horizon sec}
computes the sound horizon $r_s$, which
is the maximum size of an overdense 
fluctuation
at the time of decoupling, and 
the angle $ \theta_s = r_s/r_d $
subtended by it in the CMB\@.
This calculation relates 
the first peak in the TT spectrum
of Fig.~\ref{CMB fit}
to the density of dark energy,
the Hubble constant, and the age
of the universe.
Finally in
section~\ref{Sensitivity of the Sound Horizon to Lambda sec}
it is shown that the 
angle $\theta_s = r_s/r_d$,
which the Planck data determine as
$ \theta_s = 0.0104101 \pm 0.0000029 $,
varies by a factor of 146 
when the Hubble constant is held fixed 
but the cosmological constant $\Lambda$
is allowed to vary from zero to 
twice the value determined by
the Planck collaboration. 
This variation is mainly due to
that of $r_d$ not $r_s$.
\par
The paper does not discuss 
how the CMB anisotropies 
may have arisen from 
fluctuations in quantum fields
before or during the big bang;
this very technical subject
is sketched by 
Guth~\citep{Guth:2003rn, *Guth:2004tw, *Guth:2013epa}
and described by
Mukhanov, Feldman, and 
Brandenberger~\citep{Mukhanov:1990me, Mukhanov:2005sc}
and by Liddle and
Lyth~\citep{Liddle:1993fq}\@.

\section{The Standard Model of Cosmology
\label{The Standard Model of Cosmology sec}}

On large scales, 
our universe is homogeneous
and isotropic.
A universe in which space is maximally
symmetric~\citep[sec.\,13.24]{CahillCUP2} 
is described by a 
Friedmann-Lema{\^i}tre-Robinson-Walker 
(FLRW) universe in which 
the invariant
squared separation between two nearby 
points is
\begin{equation}
ds^2 = {} g_{ik} \, dx^i dx^k
= - c^2 dt^2 + a^2(t) \,
\left( \frac{dr^2}{1 - k \, r^2/L^2} + 
r^2 \, d\theta^2 + r^2 \, \sin^2\theta \, d\phi^2
\right)
\label {cosmo 1}
\end{equation}
\citep{FriedmannA1922, LemaitreG1927,
RobertsonHP1935,
WalkerAG1936}, 
\citep[sec.\,13.42] {CahillCUP2}.
In this model,
space (but not time) expands 
with a scale factor $a(t)$
that depends
on time but not on position.
Space is flat and infinite if $k=0$,
spherically curved and finite if $k=1$,
and hyperbolically curved and infinite 
if $k = {} -1$.
The curvature length
$L$ lets us measure our 
\tbf{comoving}
radial coordinate $r$ in meters.
\par
Einstein's equations imply
Friedmann's equation for the
Hubble expansion rate $H = \dot a /a$
\begin{equation}
H^2 = \lt(\frac{\dot a}{a}\rt)^2
={} \frac{8 \pi G}{3}  
\rho \, - \, \frac{c^2 k}{a^2 L^2} 
\label {Friedmann's equation}
\end{equation}
in which $\rho$
is a mass density that
depends on 
the scale factor $a(t)$, 
and the constant
$G={} 6.6743 \by 10^{-11}$ 
m$^3$ kg$^{-1}$ s$^{-2}$
is Newton's~\citep{Codata2018}.
The present value $H_0$
of the Hubble rate 
is the \tbf{Hubble constant} 
\begin{equation}
H_0 = \ovl H_0 \,\, h 
= 100 \,\,h 
\,\,\, \mbox{km/s/Mpc}
= \lt( 3.24078
\,\, \by 10^{-18} 
\,\, \mbox{s}^{-1} \rt) \,\, h.
\label{Hubble constant}
\end{equation}
in which $h$ (not Planck's constant)
lies in the interval
$0.67 \lesssim h \lesssim 0.74$ 
according to recent 
estimates~\citep{Aghanim:2018eyx, Riess:2019cxk, Freedman:2019jwv}.
A million parsecs (Mpc) is
$3.2616$ million lightyears (ly).
\par
The \tbf{critical density}
$\rho_{c}$ is the \tbf{flat space density}
\begin{equation}
\rho_{c} ={} \frac{3 H^2}{8 \pi G}
\end{equation}
which satisfies Friedmann's
equation (\ref{Friedmann's equation})
in flat ($k=0$) space
\begin{equation}
H^2 = \lt(\frac{\dot a}{a}\rt)^2
={} \frac{8 \pi G}{3} \, \rho_c .
\label {flat-space Friedmann equation}
\end{equation}
The present value of the
critical density is
\begin{equation}
\rho_{c0} ={} 
\frac{3 H^2_0}{8 \pi G}
= \frac{3 \ovl H^2_0}{8 \pi G} \, h^2 
= \lt( 1.87834 \by 10^{-26} \,\,
\mbox{kg/m}^3 \rt) \,\, h^2 .
\label{formula for rho critical}
\end{equation} 
The present 
mass densities 
(\ref{the six parameters})
of baryons $\rho_{b0}$ and 
of cold dark (invisible) matter $\rho_{d0}$
divided by the present value of the
critical density $\rho_{c0}$
are the dimensionless ratios
\begin{equation}
\Omega_b = \frac{\rho_{b0}}{\rho_{c0}}
\qand
\Omega_d = \frac{\rho_{d0}}{\rho_{c0}}
\label{defines Omegas}
\end{equation}
in which the factor $h$ cancels. 
The Planck collaboration's values for 
these ratios are~\citep{Aghanim:2018eyx}  
\begin{equation}
\Omega_b \, h^2 = 0.02242 \pm 0.00014 
\qand
\Omega_d \, h^2 = 0.11933 \pm 0.00091
\label{Planck densities}
\end{equation}
in terms of which
$\rho_{b0} ={} [3 \ovl H_0^2/(8 \pi G)] 
\,\, \Omega_b \, h^2$
and 
$\rho_{d0} ={} [3 \ovl H_0^2 /(8 \pi G)]
\,\, \Omega_d \, h^2$.
The ratio of invisible matter
to ordinary matter is
$\Omega_d / \Omega_b = 5.3$.
The ratio for the 
combined mass density of
baryons and dark matter is
\begin{equation}
\Omega_{bd}\, h^2 ={} {} \Omega_b \, h^2
+ \Omega_d \, h^2
= 0.14175.
\label{Omegabd}
\end{equation}
\par
The present density of radiation
is determined by
the present temperature
$ T_0 = 2.7255 \pm 0.0006 $
\,K~\citep{Fixsen2009}
of the CMB and by
Planck's formula~\citep[ex.~5.14]{CahillCUP2} 
for the mass density of photons 
\begin{equation}
\rho_{\gamma 0} ={} \frac{\pi^2 
\left (k_B T_0\right)^4}{15 \, \hbar^3 c^5}
= 4.645086 \by 10^{-31} \,\,\mbox{kg m}^{-3} .
\label {rhophotons}
\end{equation}
Adding in three kinds of massless
Dirac neutrinos at 
$ T_{0 \nu} = (4/11)^{1/3} \, T_0 $,
we get for
the present density of
massless and nearly massless 
particles~\cite[sec.~2.1]{Weinberg2010} 
\begin{equation}
\rho_{r 0} = \left [1 + 3\left (\frac{7}{8}\right)
\left (\frac{4}{11}\right)^{4/3} \right]
\,\, \rho_\gamma = 7.809885 \by 10^{-31} \,\,\mbox{kg m}^{-3} .
\label {rhor 3}
\end{equation}
Thus the density ratio $\Omega_r \, h^2$ 
for radiation is
\begin{equation}
\Omega_r \, h^2 ={} \frac{\rho_{r 0}}
{\rho_{c0}} \, h^2
= 4.15787 \by 10^{-5}.
\end{equation}
\par
If $k=0$, then space is flat and
Friedmann's equation 
(\ref{flat-space Friedmann equation})
requires the density $\rho$ 
to always be the same as the critical density
$\rho = \rho_c = 3 H^2/(8 \pi G)$\@.
The quantity $\Omega$ is 
the present density $\rho_0$
divided by the 
present critical density $\rho_{c0}$;
in a $k=0$, spatially flat universe
it is unity
\begin{equation}
\Omega \equiv \frac{\rho_0}{\rho_{c0}} 
= 1.
\label{Omega = 1}
\end{equation}
\par
The present density of 
baryons and dark matter
$\rho_{bd}$
and that of radiation $\rho_{r0}$
do not add up to the 
critical density $\rho_{c0}$.
In our $k=0$ universe,
the difference 
is the density of empty 
space 
$\rho_\Lambda = {}
\rho_{c 0} - \rho_{bd0} - \rho_{r0}$ 
\begin{equation}
\rho_\Lambda = {} 
\rho_{c 0} - \rho_{b0} 
- \rho_{d0} - \rho_{r0}.
\end{equation}
Michael Turner called it
dark energy.
It is represented by a cosmological
constant  
$\Lambda = {} 
8 \pi G \rho_\Lambda $~\citep{Riess:1998cb, Perlmutter:1998np}\@.
\par
Any departure from $k=0$
would imply a nonzero value
for the curvature density
$\rho_k = - 3 c^2 k/(8 \pi G a^2L^2)$
and for the
dimensionless ratio
\begin{equation}
\Omega_k = \frac{\rho_{k 0}}{\rho_{c 0}}
= -\frac{c^2 k}{a^2_0 H^2_0 L^2}.
\label{Omega_k}
\end{equation}
The WiggleZ dark-energy 
survey~\citep{Blake:2011en} 
used baryon acoustic oscillations to 
estimate this ratio as
$\Omega_k ={} - 0.004 \pm 0.006$;
the WMAP~\citep{Peiris:2003ff, Bennett:2012zja} 
collaboration found it to be
$\Omega_k ={} - 0.0027 \pm 0.0039$,
consistent with zero,
and the Planck 
collaboration~\citep{Aghanim:2018eyx}
got the tighter bound
\begin{equation}
\Omega_k \equiv {}
= 0.0007 \pm 0.0019 .
\end{equation}
For these reasons, 
the base model of the 
Planck collaboration
has $k=0$, and I will use that value in 
Sections~\ref{How the Scale Factor Evolves sec}
and \ref{The Sound Horizon sec}.
\par
In flat space, 
time is represented by the real line
and space by a 3-dimensional 
euclidian space that expands with a
scale factor $a(t)$\@.
In terms of \tbf{comoving}
spherical and
rectangular coordinates,
the line element is
\begin{equation}
ds^2 ={} - c^2 dt^2 + a^2(t) \lt( 
dr^2 + r^2 d\Omega^2 \rt)
= - c^2 dt^2 + a^2(t) \lt( 
dx^2 + dy^2 + dz^2 \rt)
\label{flat-space invariant}
\end{equation}
where
$d\Omega^2 ={} 
d\theta^2 + \sin^2\theta \, d\phi^2$\@.
\par
If light goes between
two nearby points $\bos{r}$
and $\bos r'$ in empty space
in time $dt$, 
then the physical distance 
between the points is $c \, dt$.
The flat-space invariant 
(\ref{flat-space invariant})
gives that physical or proper
distance as
$c \, dt ={} a(t) \sqrt{(\bos{r - r'})^2}$.
The corresponding comoving 
distance is $\sqrt{(\bos{r - r'})^2}$.
\par
Astronomers use
coordinates in which the scale factor
at the present time $t_0$ is unity
$a(t_0) = a_0= 1$\@.
In these coordinates,
physical or proper distances 
at the present time $t_0$
are the same as comoving distances,
$a(t_0) \sqrt{(\bos{r - r'})^2} = 
\sqrt{(\bos{r - r'})^2}$\@.
\par
A photon that is
emitted at the time $t_d$ 
of decoupling 
at comoving coordinate $r_d$ 
and that comes to us through
empty space along a path
of constant $\theta, \phi$
has $ds^2 =0$, and so the formula
(\ref{flat-space invariant}) for $ds^2$
gives
\begin{equation}
c \, dt = {} - a(t)  \, dr .
\label{dt vs dr}
\end{equation}
If our comoving coordinates 
now on Earth are
$t_0$ and $r=0$, then 
the comoving radial coordinate
of the emitting atom is 
\begin{equation}
r_d ={} \int_0^{r_d} dr
= {} \int_{t_d}^{t_0} \frac{c \, dt}{a(t)} .
\label{int dt vs dr}
\end{equation}
\par
The angle $\theta$
subtended by a comoving
distance $\sqrt{(\bos{r - r'})^2}$
that is perpendicular to the line
of sight from the position
$\bos r_d$ of the emitter
to that of an observer 
now on Earth is
\begin{equation}
\theta = {} \frac{\sqrt{(\bos{r - r'})^2}}
{r_d} .
\end{equation}
\par
To find the 
distance $r_d$, we need to
know how the scale factor $a(t)$
varies with the time $t$\@.

\section{How the Scale Factor Evolves
\label{How the Scale Factor Evolves sec}}

This section begins with a discussion
of how the densities of massive particles,
of massless particles, and of dark energy
vary with the scale factor.
These densities and the assumed
flatness of space will then be used
to compute the distance from
the surface of last scattering
and the farthest distance
that will ever be observed.
\par
As space expands with the scale 
factor $a(t)$, the density
of massive particles falls as
\begin{equation}
\rho_{bd} = {}\frac{\rho_{bd0}}{a^3(t)}
= \frac{\Omega_{bd} \, \rho_{c 0}}
{a^3(t)} . 
\end{equation}
Because wavelengths
stretch with $a(t)$,
the density $\rho_r$ of 
radiation falls faster 
\begin{equation}
\rho_r ={} \frac{\rho_{r0}}{a^4(t)}
= \frac{\Omega_{r} \, \rho_{c 0}}
{a^4(t)}.
\end{equation}
The density of empty space
$\rho_\Lambda$ does not
vary with the scale factor
\begin{equation}
\rho_\Lambda ={}
\Omega_\Lambda \, \rho_{c 0} .
\end{equation}
The Planck values for
the density ratios are
\begin{equation}
\Omega_{bd} \, h^2 = 0.14175, \quad
\Omega_{r} \, h^2 = 
4.15787 \by 10^{-5},
\qand 
\Omega_\Lambda \, h^2 =
0.31537.
\label{densities Planck}
\end{equation}
The last four equations
let us estimate when the 
density of matter $\rho_{bd}$
first equaled that of radiation 
$\rho_{bd} = \rho_r $
as when 
\begin{equation}
z+1 = \frac{1}{a} = {}
\frac{\Omega_{bd}}{\Omega_{r}}
= 3409
\end{equation}
and when the
density of dark energy $\rho_\Lambda$
first equaled that of matter
$\rho_\Lambda = \rho_{bd}$
as when
\begin{equation}
z+1 = \frac{1}{a} = {} \lt( 
\frac{\Omega_\Lambda{}}{\Omega_{bd}}
\rt)^{1/3} = 1.305.
\end{equation}
To relate the red shift $z$ and 
the scale factor $a=1/(z+1)$
to the time since the moment
of infinite redshift,
we need to how the scale factor
changes with time.
\par
Friedmann's equation for flat space 
(\ref{flat-space Friedmann equation})
and the formula 
(\ref{formula for rho critical})
for the critical density $\rho_{c 0}$
give the square of the 
Hubble rate as
\begin{equation}
\begin{split}
H^2 ={} & \frac{8 \pi G}{3} \, \rho
= \frac{8 \pi G}{3} \, \lt(
\rho_{bd} +\rho_r + \rho_\Lambda \rt)
= \frac{8 \pi G}{3} \, \lt(
\frac{\Omega_{bd}}{a^3} 
+\frac{\Omega_r }{a^4}
+ \Omega_\Lambda \rt) \, \rho_{c 0}
\\
={}&
\frac{8 \pi G}{3} \, \lt(
\frac{\Omega_{bd}}{a^3} 
+\frac{\Omega_r }{a^4}
+ \Omega_\Lambda \rt) 
\frac{3 \ovl H_0^2}{8 \pi G} \, h^2
= \ovl H_0^2 \,
\lt( \frac{\Omega_{bd} \, h^2}{a^3} 
+\frac{\Omega_r  \, h^2}{a^4}
+ \Omega_\Lambda \, h^2 \rt). 
\label{H in terms of densities}
\end{split}
\end{equation}
This equation evaluated 
at the present time $t_0$
at which $H=H_0$ and $a(t_0)=1$ is
\begin{equation}
H_0^2 ={} 
\ovl H_0^2 \,
\lt( \Omega_{bd} \, h^2
+ \Omega_r  \, h^2
+ \Omega_\Lambda \, h^2 \rt)
= H_0^2 \,
\lt( \Omega_{bd} 
+ \Omega_r  
+ \Omega_{\Lambda} \rt) 
\end{equation}
which restates the flat-space relation
(\ref{Omega = 1})
\begin{equation}
\Omega = \Omega_{bd} 
+ \Omega_r  
+ \Omega_{\Lambda} = 1.
\end{equation}
\par
\begin{figure}[htbp]
\begin{center}
\tbs{Redshift and scalefactor over last 14 Gyr}
\par\vspace{0.001in}
\includegraphics[width=5in, trim={0in 0in 0in 0.3in}, clip]{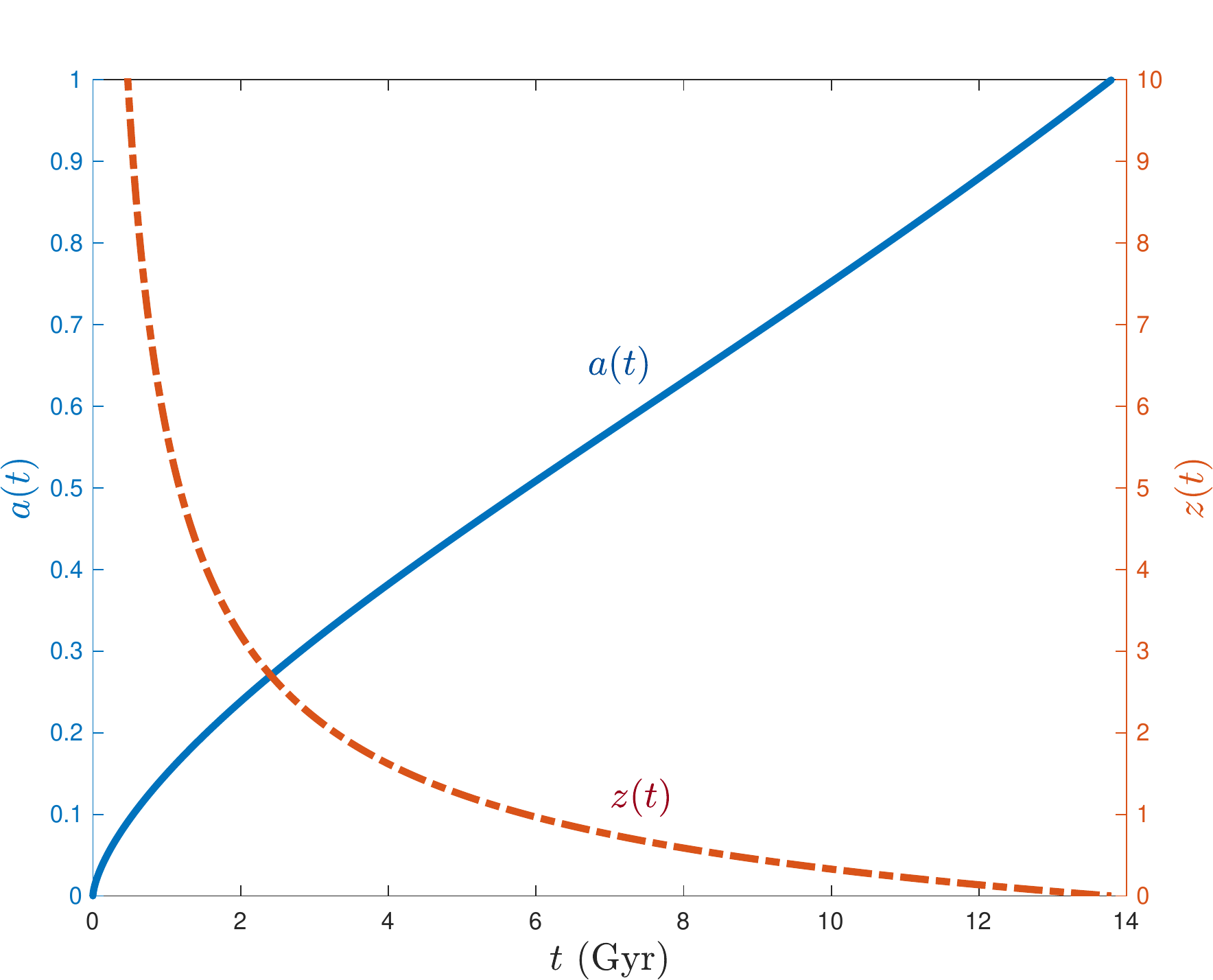}
\caption{The scale factor $ a(t) $ (solid, blue, left axis)
and redshift $ z(t) $ (dotdash, red, right axis)
are plotted against the time in Gyr
(Fig.~13.2 of \citep{CahillCUP2}, reprinted with permission).}
\label {14Gy fig}
\end{center}
\end{figure}

Using the formula 
(\ref{H in terms of densities})
for $H$ and a little calculus
\begin{equation}
dt = \frac{da}{\dot a}
= \frac{da}{a (\dot a/a)}
= \frac{da}{a H},
\label{dt/a=}
\end{equation}
we find as the time $t$ 
elapsed since $t=0$ when
the scale factor was zero
$a(0) = 0$ as
\begin{equation}
t ={} 
\int_{0}^{t} 
dt
={}
\int_{0}^{a(t)} 
\frac{da}{a H}
={}
\frac{1}{\ovl H_0}
\int_{0}^{a(t)} 
\frac{da}{\sqrt{
\Omega_\Lambda \, h^2 \, a^2
+ \Omega_{bd} \, h^2 / a  
+ \Omega_{r} \, h^2/a^2} } 
\label{the distance integral}
\end{equation}
in which the definition (\ref{Hubble constant})
of $\ovl H_0$
is $\ovl H_0 = 100$ km/s/Mpc
$ = 3.24078 \,\, \by 10^{-18}$ s$^{-1}$.
Numerical integration leads to the values 
$z(t)$ and $a(t)$ plotted 
in Fig.~\ref{14Gy fig}.

\par
Again using the formula 
(\ref{H in terms of densities})
for $H$ and a little calculus
\begin{equation}
\frac{dt}{a} = \frac{da}{a \dot a}
= \frac{da}{a^2 (\dot a/a)}
= \frac{da}{a^2 H},
\label{dt/a=}
\end{equation}
we find as
the comoving distance 
travelled by a radially moving
photon between $t_1$ and $t_2$ 
\begin{equation}
r_2 - r_1 ={} 
\int_{t_1}^{t_2} 
\frac{c \, dt}{a(t)}
={}
\int_{a(t_1)}^{a(t_2)} 
\frac{c \, da}{a^2 H}
={}
\frac{c}{\ovl H_0}
\int_{a(t_1)}^{a(t_2)} 
\frac{da}{\sqrt{
\Omega_\Lambda \, h^2 \, a^4
+ \Omega_{bd} \, h^2 \, a  
+ \Omega_{r} \, h^2} } .
\label{the distance integral}
\end{equation}
Thus the comoving distance $r_d$
from the surface of last scattering
at the time $t_d$ of decoupling
at $a(t_d) = 1/1091$ to
$r=0$ at $t_0$ is
\begin{equation}
r_d ={} 
\int_{t_d}^{t_0} 
\frac{c \, dt}{a(t)}
={}
\frac{c}{\ovl H_0}
\int_{1/1091}^{1} 
\frac{da}{\sqrt{
\Omega_\Lambda \, h^2 \, a^4
+ \Omega_{bd} \, h^2 \, a  
+ \Omega_{r} \, h^2 } } .
\label{the r_d distance integral}
\end{equation}
Substituting the values 
(\ref{densities Planck}),
we get as
the distance from the surface
of last scattering
\begin{equation}
r_d ={} 4.29171\times 10^{26}
\, \mbox{m} =
1.39085\by 10^{4}
\,\, \mbox{Mpc}
= 4.53634 \by 10^{10} 
\,\, \mbox{ly}.
\label{distance from the surface}
\end{equation}
\par
At the time of decoupling,
the physical distance $a(t_d) r_d$
of the surface of last scattering
from us was
$a(t_d) r_d = r_d/1091 = 4.16 \by 10^7$ ly.
A signal traveling that distance in 
time $t_d = 3.8 \by 10^5$ y would have 
had a speed of more than 
100 $c$~\citep{Davis:2003ad}.
Yet the the CMB coming 
to us from opposite directions
is at almost
the same temperature.
Two explanations
for this paradox are:
that the hot big bang
was preceded by a short period
of superluminal expansion called 
\tbf{inflation}~\citep{Guth:1980zm, Linde:1981mu}
and 
that the universe
equilibrated while collapsing
before the hot big 
bang~\citep{Steinhardt:2001vw, Ijjas:2018qbo} ---
a \tbf{bouncing universe}.
\par
The scale factor is $a(0) =0$
at $t=0$,
the time of infinite redshift, 
and is $a(\infty) = \infty$
at $t = \infty$ 
infinitely far in the future.
Thus the comoving radial
coordinate $r_\infty$ of the 
most distant emission of a photon
that we could receive at $r=0$
if we waited
for an infinitely long time is 
given by the integral 
\begin{equation}
\begin{split}
r_\infty  ={} &
\int_{0}^{\infty} 
\frac{c \, dt}{a(t)}
={}
\frac{c}{\ovl H_0}
\int_{0}^{\infty} 
\frac{da}{\sqrt{
\Omega_\Lambda \, h^2 \, a^4
+ \Omega_{bd} \, h^2 \, a  
+ \Omega_{r}  \, h^2} }
\\
={}&
5.94739\times 10^{26}
\, \mbox{m} =
1.92742 \by 10^{4}
\,\, \mbox{Mpc}
= 6.2864 \by 10^{10} 
\, \mbox{ly}
\label{integral for r_infinity}
\end{split}
\end{equation}
which
converges because of the 
vacuum-energy term 
$\Omega_\Lambda \, a^4$
in its denominator.
Light emitted at the time 
of the big bang farther from Earth
than 63 billion lightyears 
will never reach us
because dark energy is accelerating
the expansion of the universe.

\section{The Sound Horizon
\label{The Sound Horizon sec}}

This section begins with a discussion
of how rapidly changes in density
can propagate in the hot plasma
of dark matter, baryons, electrons,
and photons before decoupling.
This sound speed 
is then used to compute the maximum
distance $r_s$ that a density fluctuation 
could propagate from the time
of the big bang to the time
of decoupling.
The resulting sound horizon $r_s$ 
is the maximum
radius of a fluctuation in the CMB.
The angle subtended by 
such a fluctuation is the 
sound horizon divided by the
distance $r_d$ 
(\ref{distance from the surface})
from the surface of last scattering,
$\theta_s = r_s/r_d$.
We will compute this angle
as well as the Hubble constant and 
the age of the universe
by using the Planck values 
for the densities of matter, 
radiation, and dark energy,
and the assumption that 
space is flat.
\par 
Before photons decoupled from
electrons and baryons, 
the oscillations of the 
plasma of dark matter, baryons,
electrons, and photons
were contests between
gravity and radiation pressure.
Because photons vastly outnumber
baryons and electrons,
the photons determined 
the speed of sound $v_s$
in the plasma.
The pressure $p$ of a gas of photons
is one-third of its energy density
$p = \rho \, c^2/3$, and 
so the  speed of sound 
due to the 
photons is~\citep[Sec.~\textsc{iii}.6]{Zee:2013dea}
\begin{equation}
v_s = \lt(\frac{\d p}{\d \rho}\rt)^{1/2}
= \frac{c}{\sqrt{3}}.
\end{equation}
A better estimate of the
speed of sound is one that takes
into account the 
baryons~\citep[ch.~2]{Weinberg2010}
\begin{equation}
v_s = \frac{c}{\sqrt{3}}
\frac{1}{\sqrt{1+R}}
\label{speed of sound}
\end{equation}
in which $R$ is proportional to the
baryon density (\ref{the six parameters})
divided by the photon density
(\ref{rhophotons})
\begin{equation}
R = \frac{3 \rho_b}{4\rho_\c}
= \frac{3 \rho_{b 0}}{4\rho_{\c 0}} \, a
= 678.435 \, a .
\label{R}
\end{equation}
\par
The sound horizon $r_s$ 
is the comoving distance
that a pressure or sound wave 
could travel between the time
of infinite redshift
and the time of decoupling.
The high-density bubble
is a sphere, so we can compute
the distance $r_s$ for constant $\theta, \phi$\@.
Using the ratios $\Omega_{bd}$
and $\Omega_r$ in the distance integral 
(\ref{the distance integral})
with lower limit $a(0) = 0$ and upper limit
$a_d = 1/1091$ and with the speed of light
$c$ replaced by the speed of sound
$v_s$ (\ref{speed of sound}
and \ref{R}), we get
\begin{equation}
\begin{split}
r_s  ={} &
\int_{0}^{t_d} 
\frac{c \, dt}{\sqrt{3(1+R)} \, a(t)}
={}
\frac{c}{\sqrt{3} \,\, \ovl H_0} 
\int^{1/1091}_0
\frac{da}{\sqrt{(1+R)
(\Omega_\Lambda \, h^2 \, a^4
+ \Omega_{bd} \, h^2 \, a  
+ \Omega_{r} \, h^2) } } .
\label{rstar}
\end{split}
\end{equation}
Substituting the values 
$\Omega_{bd} \, h^2 = 0.14175$,
$\Omega_{r} \, h^2 = 
4.15787 \by 10^{-5}$,
and
$\Omega_\Lambda \, h^2 =
0.31537$, we find for
the sound horizon
\begin{equation}
r_s ={} 4.4685\times 10^{24}
\, \mbox{m}
=  144.814 \, \mbox{Mpc}
= 4.72321\by 10^8 \,\, 
\mbox{ly}.
\end{equation}
\par
The angle subtended by
the sound horizon $r_s$ 
at the distance $r_d$
is the ratio
\begin{equation}
\theta_s ={}
\frac{r_s}{r_d}
= 0.0104119
\end{equation}
which is exactly the
Planck result
$\theta_P ={} 0.0104119 \pm
0.0000029$, 
a measurement with 
a precision of 
0.03 \%~\citep{Aghanim:2018eyx}.
It is the location
of the first peak in the
TT spectrum of Fig.~\ref{CMB fit}.
Had we done this calculation
(\ref{the r_d distance integral} and
\ref{rstar})
of the angle $\theta_s$ for a variety
of values for the dark-energy density
$\Omega_\Lambda h^2$,
we would have found
$\Omega_\Lambda h^2 = 0.31537$
as the best density.
Thus the Planck measurement
of $\theta_s$ together with 
the flatness of space and
the densities of matter and
radiation determine the 
density of dark energy.
\par
The formula 
(\ref{H in terms of densities})
for the Hubble rate in terms of the
densities gives the Hubble constant as
\begin{equation}
H_0 ={}  \ovl H_0
\sqrt{ \Omega_\Lambda \, h^2
+ \Omega_{bd} \, h^2
+ \Omega_r \, h^2}  
={} 67.614 \,\,
\mbox{km s}^{-1}\,\mbox{Mpc}^{-1}
= 2.19121 \by 10^{-18} \,\,\mbox{s}^{-1}.
\end{equation}
This value is well within $1\s$ of the value
found by the Planck collaboration
$H_0 ={} 67.66 \pm 0.42$ 
km/s/Mpc~\citep{Aghanim:2018eyx}.
The Planck value for $H_0$
reflects the physics 
of the universe between
the big bang and decoupling
some 380,000 years later.
Using the Hubble Space Telescope 
to observe 70 long-period Cepheids
in the Large Magellanic Cloud,
the Riess group recently
found~\citep{Riess:2019cxk}
$H_{0R} = {} 74.03 \pm 1.42$ 
km/s/Mpc, a value that
reflects the physics
of the present universe.
More recently, 
using a calibration of the 
tip of the red giant branch
applied to Type Ia supernovas,
the Freedman group
found~\citep{Freedman:2019jwv}
$H_{0F} = {} 69.8 \pm 0.8$ 
km/s/Mpc, another value that
reflects the physics
of the present universe.
\par
Finally, using the formula  
(\ref{H in terms of densities})
for the Hubble rate $H=\dot a / a$
in terms of the density ratios,
we can write 
the age of the universe
as an integral of
$dt = da/(a H)$ 
from $a(0) =0$ to $a(t_0) = 1$
\begin{align}
t_0 ={}& \int_0^1 \frac{da}{a H}
= \int_0^1 \frac{da}
{\ovl H_0 
\sqrt{\Omega_\Lambda \, h^2 \, a^2
+ \Omega_{bd} \, h^2 \, a^{-1}  
+ \Omega_{r} \, h^2 \, a^{-2}}}
\\
={}&
4.35756 \by 10^{17} \, \mbox{s}
= 13.808 \by 10^9 \,\, \mbox{sidereal years},
\end{align}
which is within $1\s$ of 
the Planck collaboration's
value $t_0 = (13.787 \pm 0.020) \by 10^9$ y~\citep{Aghanim:2018eyx}.

\section{Sensitivity of the Sound Horizon to $\Lambda$
\label{Sensitivity of the Sound Horizon to Lambda sec}}

In this section,
keeping the Hubble constant fixed 
but relaxing the assumption that space is flat,
we will compute the distances
$r_s$ and $r_d$ and the angle
$\theta_s = r_s/r_d$ for different
values of the dark-energy density.
We will find that although
the sound horizon $r_s$ remains
fixed, the distance $r_d$ and the angle
$\theta_s$ vary markedly as
the cosmological constant runs
from zero to twice the Planck value.
This wide variation supports the
conclusion that space is flat and
that the dark-energy density
is close to the Planck value.
\par
Since astronomical observations
have determined the value
of the Hubble constant $H_0$
to within 10\%,
we will keep it fixed at the value estimated
by the Planck collaboration
while varying the density of
dark energy
and seeing how that shifts
the position $\theta_s$
of the first peak 
of the TT spectrum  
of Fig.~\ref{CMB fit}\@.
Since the energy density will not be 
equal to the critical density,
space will not be flat, so we must use
the Friedmann equation 
(\ref{Friedmann's equation})
\begin{equation}
H^2 = \lt(\frac{\dot a}{a}\rt)^2
={} \frac{8 \pi G}{3} \, 
\rho - \frac{c^2 k}{a^2 L^2} 
= \ovl H_0^2 \,
\lt( \frac{\Omega_{bd} \, h^2}{a^3} 
+\frac{\Omega_r  \, h^2}{a^4}
+ \Omega_{\Lambda_k} \, h^2 \rt)
- \frac{c^2 k}{a^2 L^2} 
\label {Friedmann's equation redux}
\end{equation}
which includes a curvature term
instead of the flat-space
Friedmann equation 
(\ref{Friedmann's equation})\@.
Since we are holding the Hubble
constant $H_0$ fixed,
the curvature term $- c^2 k/(a_0^2L^2)$ must
compensate for the change of the
cosmological-constant ratio
to $\Omega_{\Lambda_k}$ from
the Planck value 
$\Omega_{\Lambda} 
= 0.6889 \pm 0.0056 $~\citep{Aghanim:2018eyx}.
We can find the needed values
of $k$ and $L$ by using 
the equation (\ref{H in terms of densities})
for the Hubble constant.
We require
\begin{equation}
H_0^2 ={} 
\ovl H_0^2 \,
\lt( \Omega_{bd} \, h^2
+ \Omega_r  \, h^2
+ \Omega_{\Lambda} \, h^2 \rt)
= \ovl H_0^2 \,
\lt( \Omega_{bd} \, h^2
+ \Omega_r  \, h^2
+ \Omega_{\Lambda_k} \, h^2 \rt)
- \frac{c^2 k}{a^2_0 L^2}
\end{equation}
or
\begin{equation}
 \frac{c^2 k}{a^2_0 L^2} ={}
\ovl H_0^2 \, h^2 
\lt( \Omega_{\Lambda_k}
- \Omega_{\Lambda} \rt)
\qand
L = {}
\frac{c}{a_0H_0 |\Omega_{\Lambda_k}
- \Omega_{\Lambda}|}
\end{equation}
in which 
$k = (\Omega_{\Lambda_k} - \Omega_\Lambda)
/|\Omega_{\Lambda_k} - \Omega_\Lambda|
= \pm 1 $.
With these values of $k$ and $L$, 
Friedmann's equation 
(\ref{Friedmann's equation redux})
is
\begin{equation}
\begin{split}
H^2 = {}& 
\lt(\frac{\dot a}{a}\rt)^2
={} \ovl H^2_0 \,
\lt(  \frac{\Omega_{bd} \, h^2}{a^3} 
+\frac{\Omega_r  \, h^2}{a^4}
+ \Omega_{\Lambda_k} \, h^2 \rt)
- \frac{c^2 k}{a^2 L^2}
\\
={}&
\ovl H^2_0 \,
\lt(  \frac{\Omega_{bd} \, h^2}{a^3} 
+\frac{\Omega_r  \, h^2}{a^4}
+ \Omega_{\Lambda_k} \, h^2 
\lt(1 - \frac{1}{a^2}\rt)
+ \Omega_\Lambda \, h^2\rt).
\label {Friedmann's omega k equation}
\end{split}
\end{equation}
\par
Now since $k \ne 0$
when $\Omega_{\Lambda}$ is replaced by
$\Omega_{\Lambda_k}$,
the relation (\ref{int dt vs dr})
between $c dt$ and an element 
$dr$ of the radial
comoving coordinate  
becomes the one that
follows from $ds^2=0$
when the FLRW formula
(\ref{cosmo 1}) for $ds^2$ applies
\begin{equation}
\frac{c \, dt}{a(t)} ={} 
\pm \frac{dr}{\sqrt{1 - k r^2/L^2}}.
\label{dt vs dr FLRW}
\end{equation}
The minus sign is used
for a photon emitted during 
decoupling at $t=t_d$ 
on the surface of last
scattering at $r=r_d$
and absorbed here at $r=0$
and $t=t_0$.
Integrating, we get as the 
scaled distance $D(t_0, t_d) $ 
traveled by a photon emitted 
at comoving coordinate $r_d$ 
at time $t_d$ and observed
at $r=0$ at time $t_0$ 
\begin{equation}
D(t_d, t_0) \equiv
\int_{t_d}^{t_0} \frac{c \, dt}{a(t)}
={}
\int_0^{r_d} 
\frac{dr}{\sqrt{1 - k r^2/L^2}}
={}
\lt\{
\begin{array}{ll}
L \arcsinh\big(r_d/L\big) 
& \quad \mbox{if } k = - 1 \\
r_d  & \quad \mbox{if } k = 0 \\
L\arcsin\big(r_d/L\big)
& \quad \mbox{if } k = 1
\end{array} \rt. .
\label{scaled distance}
\end{equation}
Inverting these formulas 
(\ref{scaled distance}), 
we find the comoving coordinate $r_d$
of emission 
\begin{equation}
r_d ={} 
\lt\{
\begin{array}{ll}
L \sinh\big(D(t_d, t_0)/L \big) 
 & \quad \mbox{if } k = -1 \\
D(t_d, t_0) & \quad \mbox{if } k = 0 \\
L \sin\big(D(t_d, t_0)/L\big) 
& \quad \mbox{if } k = 1
\end{array} \rt. .
\label{comoving coordinates}
\end{equation}
Using again the relation
(\ref{dt/a=}) $dt/a=da/(a^2H)$
and the formula
(\ref{Friedmann's omega k equation})
for the Hubble rate, we find
as the scaled distance 
(\ref{scaled distance})
traveled
by a photon from the
surface of last scattering
at the time of decoupling 
\begin{equation}
\begin{split}
D(t_d, t_0) ={}& 
\int_{t_d}^{t_0} 
\frac{c \, dt}{a(t)}
={}
\int_{1/1091}^{1} 
\frac{da}{a^2 H}
\\
={}&
\frac{c}{\ovl H_0}
\int_{1/1091}^{1} 
\frac{da}{\sqrt{
\Omega_{bd} \, h^2 \, a 
+\Omega_r  \, h^2
+ \Omega_{\Lambda_k} \, h^2 
\lt(a^4 - a^2\rt)
+ \Omega_\Lambda \, h^2 \, a^4} }  .
\label{integral for rd k nonzero}
\end{split}
\end{equation}

\begin{table}[h!]
\centering
\begin{tabular}{ |p{1cm}||p{1cm}|p{2cm}|p{2cm}| p{2cm}|p{2cm}|  }
 \hline
 \multicolumn{6}{|c|}{Positions of the first peak for various 
 cosmological constants $\Lambda_k$} \\
 \hline
 $ \Lambda_k$ & $k$ & $r_s$ (Mpc) & $r_d$ (Gpc) & angle $\theta_s$ &multipole $\ell$\\
 \hline
 $ 2 \Lambda$ & $1$ & 145 & 1.53   & 0.0947  & 25 \\
 $ 3 \Lambda/2$ & $1$ & 145 & 6.87  &  0.021 & 110 \\
 $ \Lambda$ & $0$& 145 & 13.9 &   0.0104 & 220 \\
 $ 2 \Lambda/3$ & ${} -1$& 145 & 23.2  & 0.00624 & 370 \\
 $ \Lambda/2$ & ${} -1$& 145 & 27.9  & 0.00518 & 450 \\ 
 $ \Lambda/3$ & ${} -1$& 145 & 33.3  & 0.00435  & 530 \\
 $ \Lambda/4$ & ${} -1$& 145 & 36.2  & 0.00400  & 580 \\
 $ 0$ & ${} -1 $& 145 & 221  & 0.00065 & 3550 \\
 \hline
\end{tabular}
\caption{The values of the parameter $k$, 
of the comoving coordinates
of the sound horizon $r_s$ and
of the surface of last scattering $r_d$,
the angle $\theta_s = r_s/r_d$,
and the approximate multipole moment
of the first peak in the TT spectrum
are listed for several values of
the cosmological constant $\Lambda_k$.}
\label{table: 1}
\end{table}

\begin{figure}[h]
\centering
\tbs{Sensitivity of the first peak to the value of the cosmological constant}
\par\bigskip
\includegraphics[width=0.95\textwidth]{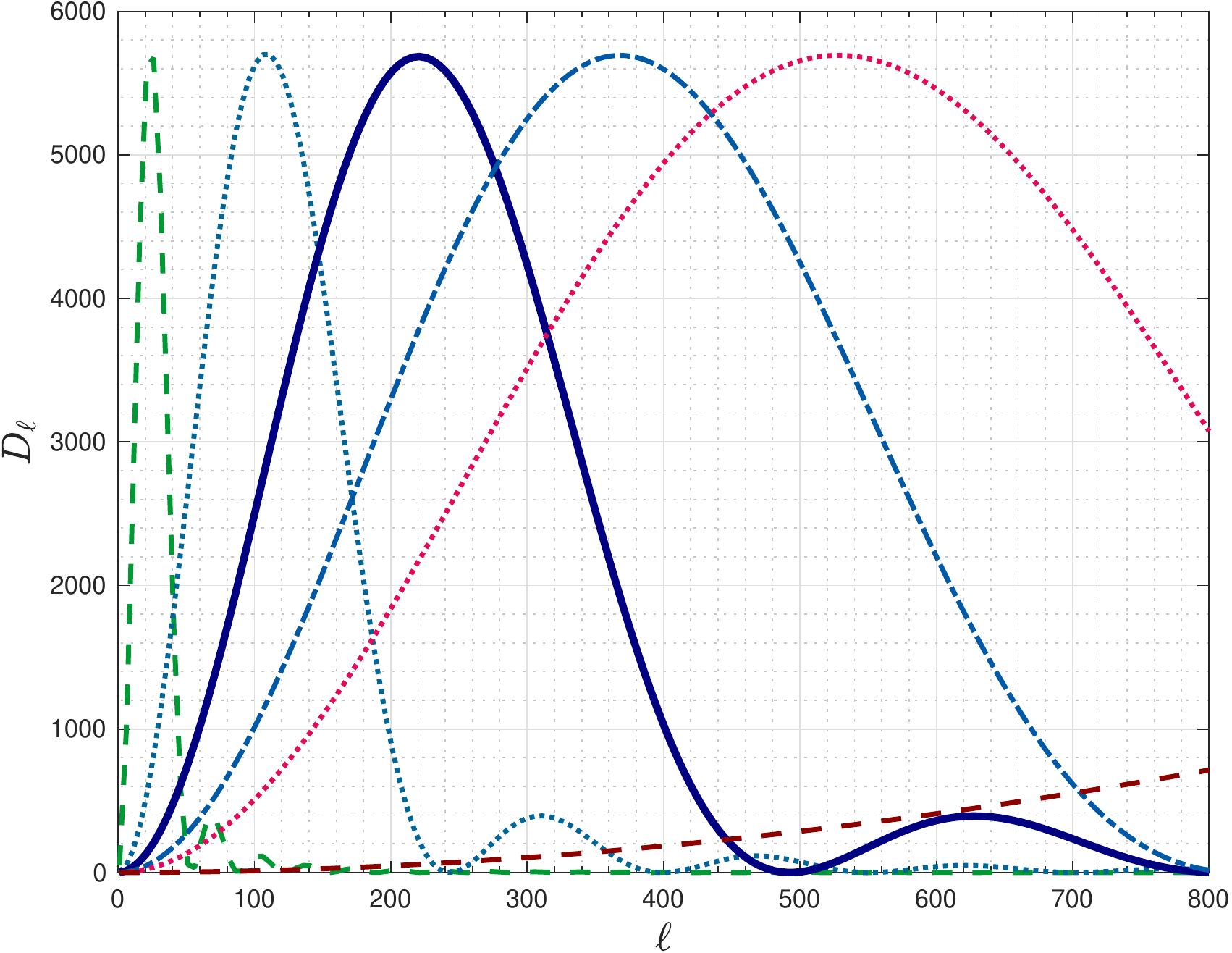}
\caption{The TT spectra of
a sky that is dark except
for a single disk at $\theta=0$ that
subtends an angle $\theta_s = r_s/r_d$ 
that is the ratio of the sound horizon $r_s$
to the distance $r_d$ from the surface of
last scattering for various values
of the cosmological constant $\Lambda_k$
with the Hubble constant $H_0$ held fixed.
From left to right, the peaks 
are at $\ell \approx 25$ and $\theta = 0.095$ 
for $\Lambda_k = 2 \Lambda$
(dashes, green),
at $\ell \approx 110$ and $\theta =0.021$
for $\Lambda_k = 3 \Lambda/2$
(dots, bluegreen),
at $\ell \approx 220$ and $\theta = 0.0104$
for $\Lambda_k = \Lambda$
(solid, dark blue),
at $\ell \approx 370$ and $\theta = 0.00624$
for $\Lambda_k = 2\Lambda/3$
(dot-dash, teal),
at $\ell \approx 530$ and $\theta = 0.00435$
for $\Lambda_k = \Lambda/3$
(dots, raspberry),
and off scale at $\ell \approx 3550$ 
and $\theta = 0.00065$ for
$\Lambda = 0$ (dashes, dark red)\@.
The spectra are scaled
so as to have the same maximum value.}
\label {one disk fit fig}
\end{figure}

\par
We use the plus sign
in the relation (\ref{dt vs dr FLRW})
between $dt$ and $dr$
for a photon emitted in the
big bang at $r=0$
and absorbed at $r_s$ at $t_d$.
So using again the relation
(\ref{dt/a=}) $dt/a=da/(a^2H)$
and the formula
(\ref{Friedmann's omega k equation})
for the Hubble rate,
we find the scaled distance
a sound wave 
could go starting at the big bang
and stopping at decoupling
would be  
\begin{align}
D(0, t_d) ={}&
\frac{c}{\ovl H_0}
\int^{1/1091}_0 \!\!\!\!\!
\frac{da}{\sqrt{3(1+R)(
\Omega_{bd} \, h^2 \, a 
+\Omega_r  \, h^2
+ \Omega_{\Lambda_k} \, h^2 
\lt(a^4 - a^2\rt)
+ \Omega_\Lambda \, h^2 \, a^4) } }.
\label{integral for rstar k nonzero}
\end {align}
The inversion formulas
(\ref{comoving coordinates}) then 
give the comoving coordinates
$r_d$ and $r_s$ and the 
angle $\theta_s = r_s/r_d$ 
corresponding to 
$D(t_0, t_d)$ and $D(t _d,0)$\@.

\par
I used these formulas
(\ref{comoving coordinates}--\ref{integral for rstar k nonzero})
to find the angles $\theta_s$
and comoving coordinates $r_s$ and 
$r_d$ that result when 
the cosmological constant 
is changed
from the Planck value $\Lambda$
to $\Lambda_k$\@.
The resulting values of 
$k, r_s, r_d, \theta_s = r_s/r_d,$
and the approximate positions of 
the first peak in the resulting TT spectrum
are listed in Table~\ref{table: 1}
for several values of the cosmological 
constant $\Lambda_k$.
The position $r_d$ of the
surface of last scattering varies
by a factor of 144, and the
angle $\theta_s$ varies 
by a factor of 146. 
The sound horizon $r_s$ is almost
independent of $\Lambda_k$
because in the integral
\ref{integral for rstar k nonzero})
for $D(t_d,0)$, the scale factor $a$
is less than $1/1091$, and
the ratios $\Omega_\Lambda$ and
$\Omega_{\Lambda_k}$ are respectively
multiplied by $a^4 \sim 10^{-12}$ and by 
$a^4 - a^2 \sim {} - 10^{-6}$.

\par
To see how the 
first peak of the TT spectrum 
might vary with the 
cosmological constant $\Lambda_k$,
I used a toy CMB consisting
of a single
disk whose radius subtends the angle
$\theta_s = r_s/r_d$.   
For such a disk about the north pole
from $\theta=0$ to $\theta=\theta_s$,
only the $m=0$ term in the formula
(\ref{alm =}) for $a_{\ell m}$
contributes, and so
\begin{equation}
a_{\ell 0} = {}
\sqrt{\pi (2 \ell +1)}
\int_{-1}^1 dx \, P_\ell(x) \, T(x)
\qand
D_\ell = \frac{\ell(\ell+1)}{2}
\lt| \int_{x_s}^1 dx \, P_\ell(x) \, T(x) \rt|^2
\end{equation}
in which $x_s = \cos \theta_s$\@.
To avoid a sharp cutoff at 
$x = \cos \theta_s$, 
I chose as the temperature distribution
across the disk 
$T(x) = {} A
\lt[1 - \cos^2( \theta_s)/x^2\rt]$
which drops smoothly to zero
across the disk. 
The resulting TT spectra 
for cosmological constants $\Lambda_k$
equal to the Planck value $\Lambda$
multiplied by 2, 3/2, 1, 2/3, 1/3, or 0
are plotted in Fig.~\ref{one disk fit fig}.
The Hubble constant was held fixed
at the Planck value.

\bibliography{physics}

\begin{thebibliography}{37}%
\makeatletter
\providecommand \@ifxundefined [1]{%
 \@ifx{#1\undefined}
}%
\providecommand \@ifnum [1]{%
 \ifnum #1\expandafter \@firstoftwo
 \else \expandafter \@secondoftwo
 \fi
}%
\providecommand \@ifx [1]{%
 \ifx #1\expandafter \@firstoftwo
 \else \expandafter \@secondoftwo
 \fi
}%
\providecommand \natexlab [1]{#1}%
\providecommand \enquote  [1]{``#1''}%
\providecommand \bibnamefont  [1]{#1}%
\providecommand \bibfnamefont [1]{#1}%
\providecommand \citenamefont [1]{#1}%
\providecommand \href@noop [0]{\@secondoftwo}%
\providecommand \href [0]{\begingroup \@sanitize@url \@href}%
\providecommand \@href[1]{\@@startlink{#1}\@@href}%
\providecommand \@@href[1]{\endgroup#1\@@endlink}%
\providecommand \@sanitize@url [0]{\catcode `\\12\catcode `\$12\catcode
  `\&12\catcode `\#12\catcode `\^12\catcode `\_12\catcode `\%12\relax}%
\providecommand \@@startlink[1]{}%
\providecommand \@@endlink[0]{}%
\providecommand \url  [0]{\begingroup\@sanitize@url \@url }%
\providecommand \@url [1]{\endgroup\@href {#1}{\urlprefix }}%
\providecommand \urlprefix  [0]{URL }%
\providecommand \Eprint [0]{\href }%
\providecommand \doibase [0]{https://doi.org/}%
\providecommand \selectlanguage [0]{\@gobble}%
\providecommand \bibinfo  [0]{\@secondoftwo}%
\providecommand \bibfield  [0]{\@secondoftwo}%
\providecommand \translation [1]{[#1]}%
\providecommand \BibitemOpen [0]{}%
\providecommand \bibitemStop [0]{}%
\providecommand \bibitemNoStop [0]{.\EOS\space}%
\providecommand \EOS [0]{\spacefactor3000\relax}%
\providecommand \BibitemShut  [1]{\csname bibitem#1\endcsname}%
\let\auto@bib@innerbib\@empty
\bibitem [{\citenamefont {Gamow}(1948)}]{Gamow:1948oesg}%
  \BibitemOpen
  \bibfield  {author} {\bibinfo {author} {\bibfnamefont {G.}~\bibnamefont
  {Gamow}},\ }\bibfield  {title} {\bibinfo {title} {{The Origin of Elements and
  the Separation of Galaxies}},\ }\href@noop {} {\bibfield  {journal} {\bibinfo
   {journal} {Phys. Rev.}\ }\textbf {\bibinfo {volume} {74}},\ \bibinfo {pages}
  {505} (\bibinfo {year} {1948})}\BibitemShut {NoStop}%
\bibitem [{\citenamefont {Alpher}\ and\ \citenamefont
  {Herman}(1950)}]{Alpher:1950zz}%
  \BibitemOpen
  \bibfield  {author} {\bibinfo {author} {\bibfnamefont {R.~A.}\ \bibnamefont
  {Alpher}}\ and\ \bibinfo {author} {\bibfnamefont {R.~C.}\ \bibnamefont
  {Herman}},\ }\bibfield  {title} {\bibinfo {title} {{Theory of the Origin and
  Relative Abundance Distribution of the Elements}},\ }\href
  {https://doi.org/10.1103/RevModPhys.22.153} {\bibfield  {journal} {\bibinfo
  {journal} {Rev. Mod. Phys.}\ }\textbf {\bibinfo {volume} {22}},\ \bibinfo
  {pages} {153} (\bibinfo {year} {1950})}\BibitemShut {NoStop}%
\bibitem [{\citenamefont {Penzias}\ and\ \citenamefont
  {Wilson}(1965)}]{Penzias:1965wn}%
  \BibitemOpen
  \bibfield  {author} {\bibinfo {author} {\bibfnamefont {A.~A.}\ \bibnamefont
  {Penzias}}\ and\ \bibinfo {author} {\bibfnamefont {R.~W.}\ \bibnamefont
  {Wilson}},\ }\bibfield  {title} {\bibinfo {title} {{A Measurement of excess
  antenna temperature at 4080-Mc/s}},\ }\href {https://doi.org/10.1086/148307}
  {\bibfield  {journal} {\bibinfo  {journal} {Astrophys. J.}\ }\textbf
  {\bibinfo {volume} {142}},\ \bibinfo {pages} {419} (\bibinfo {year}
  {1965})}\BibitemShut {NoStop}%
\bibitem [{\citenamefont {Roll}\ and\ \citenamefont
  {Wilkinson}(1966)}]{Roll:1966zz}%
  \BibitemOpen
  \bibfield  {author} {\bibinfo {author} {\bibfnamefont {P.~G.}\ \bibnamefont
  {Roll}}\ and\ \bibinfo {author} {\bibfnamefont {D.~T.}\ \bibnamefont
  {Wilkinson}},\ }\bibfield  {title} {\bibinfo {title} {{Cosmic Background
  Radiation at 3.2 cm-Support for Cosmic Black-Body Radiation}},\ }\href
  {https://doi.org/10.1103/PhysRevLett.16.405} {\bibfield  {journal} {\bibinfo
  {journal} {Phys. Rev. Lett.}\ }\textbf {\bibinfo {volume} {16}},\ \bibinfo
  {pages} {405} (\bibinfo {year} {1966})}\BibitemShut {NoStop}%
\bibitem [{\citenamefont {Mather}\ \emph {et~al.}(1990)\citenamefont {Mather}
  \emph {et~al.}}]{Mather:1991pc}%
  \BibitemOpen
  \bibfield  {author} {\bibinfo {author} {\bibfnamefont {J.~C.}\ \bibnamefont
  {Mather}} \emph {et~al.},\ }\bibfield  {title} {\bibinfo {title} {{A
  Preliminary measurement of the Cosmic Microwave Background spectrum by the
  Cosmic Background Explorer (COBE) satellite}},\ }\href
  {https://doi.org/10.1086/185717} {\bibfield  {journal} {\bibinfo  {journal}
  {Astrophys. J.}\ }\textbf {\bibinfo {volume} {354}},\ \bibinfo {pages} {L37}
  (\bibinfo {year} {1990})}\BibitemShut {NoStop}%
\bibitem [{\citenamefont {Smoot}\ \emph {et~al.}(1992)\citenamefont {Smoot}
  \emph {et~al.}}]{Smoot:1992td}%
  \BibitemOpen
  \bibfield  {author} {\bibinfo {author} {\bibfnamefont {G.~F.}\ \bibnamefont
  {Smoot}} \emph {et~al.} (\bibinfo {collaboration} {COBE}),\ }\bibfield
  {title} {\bibinfo {title} {{Structure in the COBE differential microwave
  radiometer first year maps}},\ }\href {https://doi.org/10.1086/186504}
  {\bibfield  {journal} {\bibinfo  {journal} {Astrophys. J.}\ }\textbf
  {\bibinfo {volume} {396}},\ \bibinfo {pages} {L1} (\bibinfo {year}
  {1992})}\BibitemShut {NoStop}%
\bibitem [{\citenamefont {Peiris}\ \emph {et~al.}(2003)\citenamefont {Peiris}
  \emph {et~al.}}]{Peiris:2003ff}%
  \BibitemOpen
  \bibfield  {author} {\bibinfo {author} {\bibfnamefont {H.~V.}\ \bibnamefont
  {Peiris}} \emph {et~al.} (\bibinfo {collaboration} {WMAP}),\ }\bibfield
  {title} {\bibinfo {title} {{First year Wilkinson Microwave Anisotropy Probe
  (WMAP) observations: Implications for inflation}},\ }\href
  {https://doi.org/10.1086/377228} {\bibfield  {journal} {\bibinfo  {journal}
  {Astrophys. J. Suppl.}\ }\textbf {\bibinfo {volume} {148}},\ \bibinfo {pages}
  {213} (\bibinfo {year} {2003})},\ \Eprint
  {https://arxiv.org/abs/astro-ph/0302225} {arXiv:astro-ph/0302225 [astro-ph]}
  \BibitemShut {NoStop}%
\bibitem [{\citenamefont {Bennett}\ \emph {et~al.}(2013)\citenamefont {Bennett}
  \emph {et~al.}}]{Bennett:2012zja}%
  \BibitemOpen
  \bibfield  {author} {\bibinfo {author} {\bibfnamefont {C.~L.}\ \bibnamefont
  {Bennett}} \emph {et~al.} (\bibinfo {collaboration} {WMAP}),\ }\bibfield
  {title} {\bibinfo {title} {{Nine-Year Wilkinson Microwave Anisotropy Probe
  (WMAP) Observations: Final Maps and Results}},\ }\href
  {https://doi.org/10.1088/0067-0049/208/2/20} {\bibfield  {journal} {\bibinfo
  {journal} {Astrophys. J. Suppl.}\ }\textbf {\bibinfo {volume} {208}},\
  \bibinfo {pages} {20} (\bibinfo {year} {2013})},\ \Eprint
  {https://arxiv.org/abs/1212.5225} {arXiv:1212.5225 [astro-ph.CO]}
  \BibitemShut {NoStop}%
\bibitem [{\citenamefont {Ade}\ \emph {et~al.}(2014)\citenamefont {Ade} \emph
  {et~al.}}]{Ade:2013zuv}%
  \BibitemOpen
  \bibfield  {author} {\bibinfo {author} {\bibfnamefont {P.~A.~R.}\
  \bibnamefont {Ade}} \emph {et~al.} (\bibinfo {collaboration} {Planck}),\
  }\bibfield  {title} {\bibinfo {title} {{Planck 2013 results. XVI.
  Cosmological parameters}},\ }\href
  {https://doi.org/10.1051/0004-6361/201321591} {\bibfield  {journal} {\bibinfo
   {journal} {Astron. Astrophys.}\ }\textbf {\bibinfo {volume} {571}},\
  \bibinfo {pages} {A16} (\bibinfo {year} {2014})},\ \Eprint
  {https://arxiv.org/abs/1303.5076} {arXiv:1303.5076 [astro-ph.CO]}
  \BibitemShut {NoStop}%
\bibitem [{\citenamefont {Ade}\ \emph {et~al.}(2016)\citenamefont {Ade} \emph
  {et~al.}}]{Ade:2015xua}%
  \BibitemOpen
  \bibfield  {author} {\bibinfo {author} {\bibfnamefont {P.~A.~R.}\
  \bibnamefont {Ade}} \emph {et~al.} (\bibinfo {collaboration} {Planck}),\
  }\bibfield  {title} {\bibinfo {title} {{Planck 2015 results. XIII.
  Cosmological parameters}},\ }\href
  {https://doi.org/10.1051/0004-6361/201525830} {\bibfield  {journal} {\bibinfo
   {journal} {Astron. Astrophys.}\ }\textbf {\bibinfo {volume} {594}},\
  \bibinfo {pages} {A13} (\bibinfo {year} {2016})},\ \Eprint
  {https://arxiv.org/abs/1502.01589} {arXiv:1502.01589 [astro-ph.CO]}
  \BibitemShut {NoStop}%
\bibitem [{\citenamefont {Akrami}\ \emph {et~al.}(2018)\citenamefont {Akrami}
  \emph {et~al.}}]{Akrami:2018vks}%
  \BibitemOpen
  \bibfield  {author} {\bibinfo {author} {\bibfnamefont {Y.}~\bibnamefont
  {Akrami}} \emph {et~al.} (\bibinfo {collaboration} {Planck}),\ }\href@noop {}
  {\bibinfo {title} {{Planck 2018 results. I. Overview and the cosmological
  legacy of Planck}}} (\bibinfo {year} {2018}),\ \bibinfo {note} {{A}\&A
  doi.org/10.1051/0004-6361/201833880},\ \Eprint
  {https://arxiv.org/abs/1807.06205} {arXiv:1807.06205 [astro-ph.CO]}
  \BibitemShut {NoStop}%
\bibitem [{\citenamefont {Aghanim}\ \emph {et~al.}(2018)\citenamefont {Aghanim}
  \emph {et~al.}}]{Aghanim:2018eyx}%
  \BibitemOpen
  \bibfield  {author} {\bibinfo {author} {\bibfnamefont {N.}~\bibnamefont
  {Aghanim}} \emph {et~al.} (\bibinfo {collaboration} {Planck}),\ }\href@noop
  {} {\bibinfo {title} {{Planck 2018 results. VI. Cosmological parameters}}}
  (\bibinfo {year} {2018}),\ \Eprint {https://arxiv.org/abs/1807.06209}
  {arXiv:1807.06209 [astro-ph.CO]} \BibitemShut {NoStop}%
\bibitem [{\citenamefont {Fixsen}(2009)}]{Fixsen2009}%
  \BibitemOpen
  \bibfield  {author} {\bibinfo {author} {\bibfnamefont {D.~J.}\ \bibnamefont
  {Fixsen}},\ }\bibfield  {title} {\bibinfo {title} {The temperature of the
  cosmic microwave background},\ }\href@noop {} {\bibfield  {journal} {\bibinfo
   {journal} {Astrophys. J.}\ }\textbf {\bibinfo {volume} {707}},\ \bibinfo
  {pages} {916} (\bibinfo {year} {2009})},\ \Eprint
  {https://arxiv.org/abs/0911.1955 [astro-ph.CO]} {arXiv:0911.1955
  [astro-ph.CO]} \BibitemShut {NoStop}%
\bibitem [{\citenamefont {Weinberg}(2010)}]{Weinberg2010}%
  \BibitemOpen
  \bibfield  {author} {\bibinfo {author} {\bibfnamefont {S.}~\bibnamefont
  {Weinberg}},\ }\href@noop {} {\emph {\bibinfo {title} {Cosmology}}}\
  (\bibinfo  {publisher} {Oxford University Press},\ \bibinfo {year}
  {2010})\BibitemShut {NoStop}%
\bibitem [{\citenamefont {Guth}(2003)}]{Guth:2003rn}%
  \BibitemOpen
  \bibfield  {author} {\bibinfo {author} {\bibfnamefont {A.~H.}\ \bibnamefont
  {Guth}},\ }\bibfield  {title} {\bibinfo {title} {{Inflation and cosmological
  perturbations}},\ }in\ \href@noop {} {\emph {\bibinfo {booktitle} {{The
  future of theoretical physics and cosmology: Celebrating Stephen Hawking's
  60th birthday. Proceedings, Workshop and Symposium, Cambridge, UK, January
  7-10, 2002}}}}\ (\bibinfo {year} {2003})\ pp.\ \bibinfo {pages} {725--754},\
  \Eprint {https://arxiv.org/abs/astro-ph/0306275} {arXiv:astro-ph/0306275
  [astro-ph]} \BibitemShut {NoStop}%
\bibitem [{\citenamefont {Guth}(2004)}]{Guth:2004tw}%
  \BibitemOpen
  \bibfield  {author} {\bibinfo {author} {\bibfnamefont {A.~H.}\ \bibnamefont
  {Guth}},\ }\bibfield  {title} {\bibinfo {title} {{Inflation}},\ }in\
  \href@noop {} {\emph {\bibinfo {booktitle} {{Measuring and modeling the
  universe. Proceedings, Symposium, Pasadena, USA, November 17-22, 2002}}}}\
  (\bibinfo {year} {2004})\ pp.\ \bibinfo {pages} {31--52},\ \Eprint
  {https://arxiv.org/abs/astro-ph/0404546} {arXiv:astro-ph/0404546 [astro-ph]}
  \BibitemShut {NoStop}%
\bibitem [{\citenamefont {Guth}(2013)}]{Guth:2013epa}%
  \BibitemOpen
  \bibfield  {author} {\bibinfo {author} {\bibfnamefont {A.~H.}\ \bibnamefont
  {Guth}},\ }\bibfield  {title} {\bibinfo {title} {{Quantum Fluctuations in
  Cosmology and How They Lead to a Multiverse}},\ }in\ \href@noop {} {\emph
  {\bibinfo {booktitle} {{Proceedings, 25th Solvay Conference on Physics: The
  Theory of the Quantum World: Brussels, Belgium, October 19-25, 2011}}}}\
  (\bibinfo {year} {2013})\ \Eprint {https://arxiv.org/abs/1312.7340}
  {arXiv:1312.7340 [hep-th]} \BibitemShut {NoStop}%
\bibitem [{\citenamefont {Mukhanov}\ \emph {et~al.}(1992)\citenamefont
  {Mukhanov}, \citenamefont {Feldman},\ and\ \citenamefont
  {Brandenberger}}]{Mukhanov:1990me}%
  \BibitemOpen
  \bibfield  {author} {\bibinfo {author} {\bibfnamefont {V.~F.}\ \bibnamefont
  {Mukhanov}}, \bibinfo {author} {\bibfnamefont {H.~A.}\ \bibnamefont
  {Feldman}},\ and\ \bibinfo {author} {\bibfnamefont {R.~H.}\ \bibnamefont
  {Brandenberger}},\ }\bibfield  {title} {\bibinfo {title} {{Theory of
  cosmological perturbations. Part 1. Classical perturbations. Part 2. Quantum
  theory of perturbations. Part 3. Extensions}},\ }\href
  {https://doi.org/10.1016/0370-1573(92)90044-Z} {\bibfield  {journal}
  {\bibinfo  {journal} {Phys. Rept.}\ }\textbf {\bibinfo {volume} {215}},\
  \bibinfo {pages} {203} (\bibinfo {year} {1992})}\BibitemShut {NoStop}%
\bibitem [{\citenamefont {Mukhanov}(2005)}]{Mukhanov:2005sc}%
  \BibitemOpen
  \bibfield  {author} {\bibinfo {author} {\bibfnamefont {V.}~\bibnamefont
  {Mukhanov}},\ }\href
  {http://www-spires.fnal.gov/spires/find/books/www?cl=QB981.M89::2005} {\emph
  {\bibinfo {title} {{Physical Foundations of Cosmology}}}}\ (\bibinfo
  {publisher} {Cambridge University Press},\ \bibinfo {address} {Oxford},\
  \bibinfo {year} {2005})\BibitemShut {NoStop}%
\bibitem [{\citenamefont {Liddle}\ and\ \citenamefont
  {Lyth}(1993)}]{Liddle:1993fq}%
  \BibitemOpen
  \bibfield  {author} {\bibinfo {author} {\bibfnamefont {A.~R.}\ \bibnamefont
  {Liddle}}\ and\ \bibinfo {author} {\bibfnamefont {D.~H.}\ \bibnamefont
  {Lyth}},\ }\bibfield  {title} {\bibinfo {title} {{The Cold dark matter
  density perturbation}},\ }\href
  {https://doi.org/10.1016/0370-1573(93)90114-S} {\bibfield  {journal}
  {\bibinfo  {journal} {Phys. Rept.}\ }\textbf {\bibinfo {volume} {231}},\
  \bibinfo {pages} {1} (\bibinfo {year} {1993})},\ \Eprint
  {https://arxiv.org/abs/astro-ph/9303019} {arXiv:astro-ph/9303019 [astro-ph]}
  \BibitemShut {NoStop}%
\bibitem [{\citenamefont {Cahill}(2019)}]{CahillCUP2}%
  \BibitemOpen
  \bibfield  {author} {\bibinfo {author} {\bibfnamefont {K.}~\bibnamefont
  {Cahill}},\ }\href@noop {} {\emph {\bibinfo {title} {\textit{Physical
  Mathematics}}}},\ \bibinfo {edition} {2nd}\ ed.\ (\bibinfo  {publisher}
  {Cambridge University Press},\ \bibinfo {year} {2019})\BibitemShut {NoStop}%
\bibitem [{\citenamefont {Friedmann}(1922)}]{FriedmannA1922}%
  \BibitemOpen
  \bibfield  {author} {\bibinfo {author} {\bibfnamefont {A.}~\bibnamefont
  {Friedmann}},\ }\bibfield  {title} {\bibinfo {title} {{\"{U}}ber die
  kr{\"{u}}mmung des raumes},\ }\href@noop {} {\bibfield  {journal} {\bibinfo
  {journal} {Z.~Phys.}\ }\textbf {\bibinfo {volume} {10}},\ \bibinfo {pages}
  {377} (\bibinfo {year} {1922})}\BibitemShut {NoStop}%
\bibitem [{\citenamefont {Lema{\^i}tre}(1927)}]{LemaitreG1927}%
  \BibitemOpen
  \bibfield  {author} {\bibinfo {author} {\bibfnamefont {G.}~\bibnamefont
  {Lema{\^i}tre}},\ }\href@noop {} {\bibfield  {journal} {\bibinfo  {journal}
  {Ann. Soc. Sci. Brux.}\ }\textbf {\bibinfo {volume} {A47}},\ \bibinfo {pages}
  {49} (\bibinfo {year} {1927})}\BibitemShut {NoStop}%
\bibitem [{\citenamefont {Robertson}(1935)}]{RobertsonHP1935}%
  \BibitemOpen
  \bibfield  {author} {\bibinfo {author} {\bibfnamefont {H.~P.}\ \bibnamefont
  {Robertson}},\ }\href@noop {} {\bibfield  {journal} {\bibinfo  {journal} {Ap.
  J.}\ }\textbf {\bibinfo {volume} {82}},\ \bibinfo {pages} {284} (\bibinfo
  {year} {1935})}\BibitemShut {NoStop}%
\bibitem [{\citenamefont {Walker}(1936)}]{WalkerAG1936}%
  \BibitemOpen
  \bibfield  {author} {\bibinfo {author} {\bibfnamefont {A.~G.}\ \bibnamefont
  {Walker}},\ }\href@noop {} {\bibfield  {journal} {\bibinfo  {journal} {Proc.
  Lond. Math. Soc. (2)}\ }\textbf {\bibinfo {volume} {42}},\ \bibinfo {pages}
  {90} (\bibinfo {year} {1936})}\BibitemShut {NoStop}%
\bibitem [{\citenamefont {Tiesinga}\ \emph {et~al.}(2019)\citenamefont
  {Tiesinga}, \citenamefont {Mohr}, \citenamefont {Newell},\ and\ \citenamefont
  {Taylor}}]{Codata2018}%
  \BibitemOpen
  \bibfield  {author} {\bibinfo {author} {\bibfnamefont {E.}~\bibnamefont
  {Tiesinga}}, \bibinfo {author} {\bibfnamefont {P.~J.}\ \bibnamefont {Mohr}},
  \bibinfo {author} {\bibfnamefont {D.~B.}\ \bibnamefont {Newell}},\ and\
  \bibinfo {author} {\bibfnamefont {B.~N.}\ \bibnamefont {Taylor}} (\bibinfo
  {collaboration} {CODATA}),\ }\href@noop {} {\bibinfo {title} {The 2018 codata
  recommended values of the fundamental physical constants}},\ \bibinfo
  {howpublished} {physics.nist.gov/constants} (\bibinfo {year}
  {2019})\BibitemShut {NoStop}%
\bibitem [{\citenamefont {Riess}\ \emph {et~al.}(2019)\citenamefont {Riess},
  \citenamefont {Casertano}, \citenamefont {Yuan}, \citenamefont {Macri},\ and\
  \citenamefont {Scolnic}}]{Riess:2019cxk}%
  \BibitemOpen
  \bibfield  {author} {\bibinfo {author} {\bibfnamefont {A.~G.}\ \bibnamefont
  {Riess}}, \bibinfo {author} {\bibfnamefont {S.}~\bibnamefont {Casertano}},
  \bibinfo {author} {\bibfnamefont {W.}~\bibnamefont {Yuan}}, \bibinfo {author}
  {\bibfnamefont {L.~M.}\ \bibnamefont {Macri}},\ and\ \bibinfo {author}
  {\bibfnamefont {D.}~\bibnamefont {Scolnic}},\ }\bibfield  {title} {\bibinfo
  {title} {{Large Magellanic Cloud Cepheid Standards Provide a 1\% Foundation
  for the Determination of the Hubble Constant and Stronger Evidence for
  Physics Beyond LambdaCDM}},\ }\href
  {https://doi.org/10.3847/1538-4357/ab1422} {\bibfield  {journal} {\bibinfo
  {journal} {Astrophys. J.}\ }\textbf {\bibinfo {volume} {876}},\ \bibinfo
  {pages} {85} (\bibinfo {year} {2019})},\ \Eprint
  {https://arxiv.org/abs/1903.07603} {arXiv:1903.07603 [astro-ph.CO]}
  \BibitemShut {NoStop}%
\bibitem [{\citenamefont {Freedman}\ \emph {et~al.}(2019)\citenamefont
  {Freedman} \emph {et~al.}}]{Freedman:2019jwv}%
  \BibitemOpen
  \bibfield  {author} {\bibinfo {author} {\bibfnamefont {W.~L.}\ \bibnamefont
  {Freedman}} \emph {et~al.},\ }\href@noop {} {\bibinfo {title} {{The
  Carnegie-Chicago Hubble Program. VIII. An Independent Determination of the
  Hubble Constant Based on the Tip of the Red Giant Branch}}} (\bibinfo {year}
  {2019}),\ \Eprint {https://arxiv.org/abs/1907.05922} {arXiv:1907.05922
  [astro-ph.CO]} \BibitemShut {NoStop}%
\bibitem [{\citenamefont {Riess}\ \emph {et~al.}(1998)\citenamefont {Riess}
  \emph {et~al.}}]{Riess:1998cb}%
  \BibitemOpen
  \bibfield  {author} {\bibinfo {author} {\bibfnamefont {A.~G.}\ \bibnamefont
  {Riess}} \emph {et~al.} (\bibinfo {collaboration} {Supernova Search Team}),\
  }\bibfield  {title} {\bibinfo {title} {{Observational evidence from
  supernovae for an accelerating universe and a cosmological constant}},\
  }\href {https://doi.org/10.1086/300499} {\bibfield  {journal} {\bibinfo
  {journal} {Astron. J.}\ }\textbf {\bibinfo {volume} {116}},\ \bibinfo {pages}
  {1009} (\bibinfo {year} {1998})},\ \Eprint
  {https://arxiv.org/abs/astro-ph/9805201} {arXiv:astro-ph/9805201 [astro-ph]}
  \BibitemShut {NoStop}%
\bibitem [{\citenamefont {Perlmutter}\ \emph {et~al.}(1999)\citenamefont
  {Perlmutter} \emph {et~al.}}]{Perlmutter:1998np}%
  \BibitemOpen
  \bibfield  {author} {\bibinfo {author} {\bibfnamefont {S.}~\bibnamefont
  {Perlmutter}} \emph {et~al.} (\bibinfo {collaboration} {Supernova Cosmology
  Project}),\ }\bibfield  {title} {\bibinfo {title} {{Measurements of $\Omega$
  and $\Lambda$ from 42 high redshift supernovae}},\ }\href
  {https://doi.org/10.1086/307221} {\bibfield  {journal} {\bibinfo  {journal}
  {Astrophys. J.}\ }\textbf {\bibinfo {volume} {517}},\ \bibinfo {pages} {565}
  (\bibinfo {year} {1999})},\ \Eprint {https://arxiv.org/abs/astro-ph/9812133}
  {arXiv:astro-ph/9812133 [astro-ph]} \BibitemShut {NoStop}%
\bibitem [{\citenamefont {Blake}\ \emph {et~al.}(2011)\citenamefont {Blake}
  \emph {et~al.}}]{Blake:2011en}%
  \BibitemOpen
  \bibfield  {author} {\bibinfo {author} {\bibfnamefont {C.}~\bibnamefont
  {Blake}} \emph {et~al.},\ }\bibfield  {title} {\bibinfo {title} {{The WiggleZ
  Dark Energy Survey: mapping the distance-redshift relation with baryon
  acoustic oscillations}},\ }\href
  {https://doi.org/10.1111/j.1365-2966.2011.19592.x} {\bibfield  {journal}
  {\bibinfo  {journal} {Mon. Not. Roy. Astron. Soc.}\ }\textbf {\bibinfo
  {volume} {418}},\ \bibinfo {pages} {1707} (\bibinfo {year} {2011})},\ \Eprint
  {https://arxiv.org/abs/1108.2635} {arXiv:1108.2635 [astro-ph.CO]}
  \BibitemShut {NoStop}%
\bibitem [{\citenamefont {Davis}\ and\ \citenamefont
  {Lineweaver}(2003)}]{Davis:2003ad}%
  \BibitemOpen
  \bibfield  {author} {\bibinfo {author} {\bibfnamefont {T.~M.}\ \bibnamefont
  {Davis}}\ and\ \bibinfo {author} {\bibfnamefont {C.~H.}\ \bibnamefont
  {Lineweaver}},\ }\bibfield  {title} {\bibinfo {title} {{Expanding confusion:
  common misconceptions of cosmological horizons and the superluminal expansion
  of the universe}},\ }\bibfield  {journal} {\bibinfo  {journal} {Proc. Astron.
  Soc. Austral.}\ }\href {https://doi.org/10.1071/AS03040} {10.1071/AS03040}
  (\bibinfo {year} {2003}),\ \bibinfo {note} {[Publ. Astron. Soc.
  Austral.21,97(2004)]},\ \Eprint {https://arxiv.org/abs/astro-ph/0310808}
  {arXiv:astro-ph/0310808 [astro-ph]} \BibitemShut {NoStop}%
\bibitem [{\citenamefont {Guth}(1981)}]{Guth:1980zm}%
  \BibitemOpen
  \bibfield  {author} {\bibinfo {author} {\bibfnamefont {A.~H.}\ \bibnamefont
  {Guth}},\ }\bibfield  {title} {\bibinfo {title} {{Inflationary universe: A
  possible solution to the horizon and flatness problems}},\ }\href
  {https://doi.org/10.1103/PhysRevD.23.347} {\bibfield  {journal} {\bibinfo
  {journal} {Phys. Rev.}\ }\textbf {\bibinfo {volume} {D23}},\ \bibinfo {pages}
  {347} (\bibinfo {year} {1981})}\BibitemShut {NoStop}%
\bibitem [{\citenamefont {Linde}(1982)}]{Linde:1981mu}%
  \BibitemOpen
  \bibfield  {author} {\bibinfo {author} {\bibfnamefont {A.~D.}\ \bibnamefont
  {Linde}},\ }\bibfield  {title} {\bibinfo {title} {{A New Inflationary
  Universe Scenario: A Possible Solution of the Horizon, Flatness, Homogeneity,
  Isotropy and Primordial Monopole Problems}},\ }\href
  {https://doi.org/10.1016/0370-2693(82)91219-9} {\bibfield  {journal}
  {\bibinfo  {journal} {Phys. Lett.}\ }\textbf {\bibinfo {volume} {108B}},\
  \bibinfo {pages} {389} (\bibinfo {year} {1982})}\BibitemShut {NoStop}%
\bibitem [{\citenamefont {Steinhardt}\ and\ \citenamefont
  {Turok}(2002)}]{Steinhardt:2001vw}%
  \BibitemOpen
  \bibfield  {author} {\bibinfo {author} {\bibfnamefont {P.~J.}\ \bibnamefont
  {Steinhardt}}\ and\ \bibinfo {author} {\bibfnamefont {N.}~\bibnamefont
  {Turok}},\ }\bibfield  {title} {\bibinfo {title} {{A Cyclic model of the
  universe}},\ }\href {https://doi.org/10.1126/science.1070462} {\bibfield
  {journal} {\bibinfo  {journal} {Science}\ }\textbf {\bibinfo {volume}
  {296}},\ \bibinfo {pages} {1436} (\bibinfo {year} {2002})},\ \Eprint
  {https://arxiv.org/abs/hep-th/0111030} {arXiv:hep-th/0111030 [hep-th]}
  \BibitemShut {NoStop}%
\bibitem [{\citenamefont {Ijjas}\ and\ \citenamefont
  {Steinhardt}(2018)}]{Ijjas:2018qbo}%
  \BibitemOpen
  \bibfield  {author} {\bibinfo {author} {\bibfnamefont {A.}~\bibnamefont
  {Ijjas}}\ and\ \bibinfo {author} {\bibfnamefont {P.~J.}\ \bibnamefont
  {Steinhardt}},\ }\bibfield  {title} {\bibinfo {title} {{Bouncing Cosmology
  made simple}},\ }\href {https://doi.org/10.1088/1361-6382/aac482} {\bibfield
  {journal} {\bibinfo  {journal} {Class. Quant. Grav.}\ }\textbf {\bibinfo
  {volume} {35}},\ \bibinfo {pages} {135004} (\bibinfo {year} {2018})},\
  \Eprint {https://arxiv.org/abs/1803.01961} {arXiv:1803.01961 [astro-ph.CO]}
  \BibitemShut {NoStop}%
\bibitem [{\citenamefont {Zee}(2013)}]{Zee:2013dea}%
  \BibitemOpen
  \bibfield  {author} {\bibinfo {author} {\bibfnamefont {A.}~\bibnamefont
  {Zee}},\ }\href@noop {} {\emph {\bibinfo {title} {{Einstein Gravity in a
  Nutshell}}}}\ (\bibinfo  {publisher} {Princeton University Press},\ \bibinfo
  {address} {New Jersey},\ \bibinfo {year} {2013})\ p.\ \bibinfo {pages}
  {866}\BibitemShut {NoStop}%
\end{thebibliography}%
 
\end{document}